\documentclass[]{aa}
\usepackage{graphicx}
\usepackage{dirtytalk}
\usepackage{currvita}
\usepackage[T1]{fontenc}
\usepackage{txfonts}
\usepackage{hyperref}
\usepackage{xcolor}
\hypersetup{
colorlinks,
linkcolor={red!80!black},
citecolor={blue!80!black},
urlcolor={blue!80!black}
}

\newenvironment{myfont}{\fontfamily{ppl}\selectfont}{\par}
\DeclareTextFontCommand{\textmyfont}{\myfont}

\newcommand{\code}[1]{\texttt{#1}}

\def\cm3{cm$^{-3}$}
\def\kms{km\,s$^{-1}$}
\def\mms{Mm\,s$^{-1}$}
\def\msunyr{$M_{\odot}$\,yr$^{-1}$}
\def\lsun{$L_{\odot}$}
\def\rsun{$R_{\odot}$}
\def\mdot{$\dot{\rm M}$}

\def\msun{$M_{\odot}$}

\def\one{\ts {\,\sc i}}
\def\two{\ts {\,\sc ii}}
\def\three{\ts {\,\sc iii}}
\def\four{\ts {\,\sc iv}}
\def\five{\ts {\sc v}}
\def\six{\ts {\sc vi}}
\def\beq{\begin{equation}}
\def\eeq{\end{equation}}
\def\rcsm{$R_{\rm CSM}$}

\def\lesssim{\mathrel{\hbox{\rlap{\hbox{\lower4pt\hbox{$\sim$}}}\hbox{$<$}}}}
\def\gtrsim{\mathrel{\hbox{\rlap{\hbox{\lower4pt\hbox{$\sim$}}}\hbox{$>$}}}}

\def\one{{\,\sc i}}
\def\two{{\,\sc ii}}
\def\three{{\,\sc iii}}
\def\four{{\,\sc iv}}
\def\five{{\sc v}}
\def\six{{\sc vi}}

\def\voned{{\code{V1D}}}

\def\mesa{{\code{MESA}}}
\def\cmfgen{{\code{CMFGEN}}}
\def\heracles{{\code{HERACLES}}}

\def\ekin{$E_{\rm kin}$}

\def\ergs{erg\,s$^{-1}$}

\begin{document}

   \title{A modeling perspective on the diversity of red-supergiant stars exploding within circumstellar material}

   \titlerunning{Type II SNe interacting with CSM}

\author{
Luc Dessart\inst{\ref{inst1}}
\and
W. V. Jacobson-Gal\'an\inst{\ref{inst2},\ref{inst3}}
 }

\institute{
    Institut d'Astrophysique de Paris, CNRS-Sorbonne Universit\'e, 98 bis boulevard Arago, F-75014 Paris, France\label{inst1}
   \and
  Cahill Center for Astrophysics, California Institute of Technology, Pasadena, CA 91125, USA\label{inst2}
  \and
  NASA Hubble Fellow\label{inst3}
}

   \date{}

  \abstract{
 With the ever faster cadence of untargeted surveys of the sky, the supernova (SN) community will capture in the coming years a growing number of shock breakouts in red-supergiant (RSG) stars. Expecting a high frequency of breakouts within circumstellar material (CSM), we have produced an extended, regular and cubic grid of models covering from low- to high-energy explosions, compact to extended CSM, moderate- to high-density CSM. Here, we document the main results from the radiation-hydrodynamics and nonlocal thermodynamic equilibrium radiative-transfer calculations over the first 15\,d of evolution, including the bolometric and multi-band light curves and the salient features from spectra. As before, CSM interaction is found to boost the UV brightness and shorten the optical rise time if compact. Higher ionization (e.g., as seen with O\,{\sc vi}\,3820\,\AA) is obtained for more compact CSM, and is maximum for explosions in a vacuum. CSM interaction also diversifies the spectral evolution as seen in line profile morphology, with electron-scattering broadening dominating during the IIn phase. In the absence of CSM, Doppler broadening dominates immediately after shock breakout and leads to strongly blueshifted emission in lines such as He\two\,4685.70\,\AA\ or C\four\,5804.86\,\AA. This treasury of models will be used to analyze as well as predict future observations of RSG shock breakouts in CSM.
}


   \maketitle


\section{Introduction}
\label{sect_intro}

In recent years, high-cadence optical surveys of the sky have allowed the routine discovery of red-supergiant (RSG) star explosions closer to shock breakout than ever before (see, e.g., \citealt{galyam_early_2p_11}). Prompt photometric and spectroscopic follow-up have revealed the heterogeneity of these early-time properties. These include a range in UV brightness as captured by Swift \citep{irani_sn2_24,wynn_pap2_24} or the frequent detection of events with spectral lines not strongly Doppler-broadened, as expected for emission from a fast expanding photosphere, but instead emission lines with a narrow core and extended, symmetric, electron-scattering broadened wings (e.g., \citealt{yaron_13fs_17}; \citealt{zhang_18zd_20}; \citealt{terreran_20pni_22}; \citealt{jacobson_galan_20tlf_22}). Such spectral signatures have been observed in hydrogen-rich (aka, Type II), core-collapse supernovae (SNe) as early as the 80's (see, e.g., \citealt{niemela_etal_85}), leading to the creation of the new Type IIn class \citep{schlegel_iin_90}. SN\,1998S has become the prototype for this class because of its extraordinary luminosity but also its unprecedented dataset including both low-resolution and high-resolution spectra, spectropolarimetry, and early- to late-time observations (e.g., \citealt{fassia_98S_00,fassia_98S_01,leonard_98S_00,shivvers_98S_15}).

The general consensus is that such ``IIn'' signatures indicate the presence of circumstellar material (CSM) immediately at the surface of the exploding star, causing the spectroscopic ``anomaly'' \citep{chugai_98S_01,d09_94w,groh_13cu,d17_13fs,boian_groh_20} but also at the origin of a bolometric boost affecting the light curve for days or weeks \citep{moriya_rsg_csm_11,D15_2n,morozova_2l_2p_17}. More recent observations covering the earliest times immediately after shock emergence at the progenitor surface suggest that the IIn signatures persist over different durations and with ionization properties that span from high values with SN\,2013fs (e.g., lines of O\six) down to low values with SN\,1998S (e.g., lines of N\three). A compendium of such events is presented in \citet{wynn_pap2_24}. Today, observations indicate that a significant fraction (i.e., several 10\,\% ) of Type II SNe are enshrouded in some CSM \citep{bruch_csm_21,bruch_csm_22}, which warrants further study on the origin and implications of this material.

The nature of this material remains unclear today. It may arise from intense mass loss in the final years prior to core collapse, perhaps from core-rooted instabilities \citep{shiode_wave_14,fuller_rsg_17} or mass exchange in interacting binaries (e.g., \citealt{ercolino_bin_24}). But in cases where the CSM is compact and directly present at the surface of the RSG progenitor, the material may simply correspond to the fundamental structure of RSG atmospheres, characterized by a sizable mass content and finite extent \citep{d17_13fs,soker_csm_21,fuller_tsuna_rsg_24}, perhaps influenced by pulsational mass loss (e.g., \citealt{yoon_cantiello_rsg_14}). Indeed, 1D and 3D hydrodynamical simulations of RSG star envelopes reveal a complex structure and numerous instabilities not typically produced by standard stellar evolution calculations (see, e.g., \citealt{goldberg_3d_rsg_22}; \citealt{ma_rsg_24,ma_rsg_25}; \citealt{bronner_rsg_25}).

In this work, we extend our previous explorations \citep{d17_13fs,dessart_wynn_23} to produce a cubic grid that regularly samples ejecta kinetic energy, CSM density, and CSM extent. By adopting a single progenitor (i.e., a solar-metallicity 15\,\msun\ star), we no doubt fall short of the likely diversity in SNe II exploding in CSM, in particular because we miss partially stripped RSG progenitors. But, as is apparent from the results, this regular grid captures the properties not only of most of the events that have been observed, but also makes predictions beyond the current frontiers for what may be to come with future observations, in particular with the strategically scheduled survey between the Vera-Rubin Large Synoptic Survey Telescope and the Zwicky Transient Factory \citep{lsst_ztf}. This grid of models will also be used to create a library of UV light curves and spectra (Dessart et al, in prep.) in anticipation for the hoped-for, future observations with ULTRASAT \citep{ULTRASAT} and UVEX \citep{uvex}.

In the next section, we present the numerical setup for the calculations, including the choice of progenitor model, the simulations of the explosion until shock breakout, the radiation-hydrodynamics calculations of the shock breakout phase out to 15\,d, and the post-processing of these results at multiple epochs with radiative transfer. We then present the results from the radiation hydrodynamics calculations in Section~\ref{sect_rhd}. We discuss the results for the multi-band light curves in Section~\ref{sect_lc_cmfgen}. Section~\ref{sect_spec} then discusses the properties of the model spectra, in particular how H$\alpha$ can be used as a probe of the evolving dynamical properties (Section~\ref{sect_ha}), the wide range in ionization properties inferred from spectra (Section~\ref{sect_ionization}), as well as the distinct signatures between the CSM and No-CSM cases (Section~\ref{sect_csm_nocsm}). We present our conclusions in Section~\ref{sect_conc}. These simulations will be confronted to the current sample of SNe II exploding in CSM in Jacobson-Gal\'an et al. (in prep.) and subsequently uploaded at \url{https://zenodo.org/communities/snrt}.

\section{Numerical setup}
\label{sect_setup}

In this work, we perform radiation hydrodynamics calculations with the code \heracles\ (\citealt{gonzalez_heracles_07}; \citealt{vaytet_mg_11}; \citealt{D15_2n}) for a variety of configurations corresponding to Type II SN ejecta interacting with CSM. For each configuration, we post-process ``snapshots'' with the nonlocal thermodynamic equilibrium (NLTE) radiative transfer code \cmfgen\ \citep{HD12} using the Sobolev, nonmonotonic solver \citep{D15_2n}. The main asset in this approach is to retain the complex temperature, density, and (nonmonotonic) velocity structure of the interaction with the disadvantage of a nonoptimal treatment of the radiative transfer (e.g., a more simplistic treatment of lines and the coupling between gas and radiation -- see for example the comparison in \citealt{dessart_98S_25}). This limitation means that the strength of some lines is uncertain (e.g., the emission blueward of He\two\,4685.70\,\AA\ due to C\,{\sc iii}\,4647.42\,\AA\ and N\three\,4640.64\,\AA\ tends to be underestimated by the nonmonotonic solver relative to observations) and thus more weight is given in this work on the line profile morphology and the relative strength between lines (e.g., to be used as an ionization diagnostic).

The goal of this study is thus to document the evolution of the gas and radiation over time (i.e., light curves and spectra) by starting at the time when the shock crosses the progenitor surface radius $R_\ast$. We thus cover in detail the rise to peak luminosity, the peak phase, and the subsequent decline until the interaction is over or until 15\,d of physical time, whichever comes first. We limited the study to the cases of relatively short-lived interactions rather than enduring interactions lasting many weeks or months and producing bona fide Type IIn SNe. However, we mapped the parameter space more systematically than in our previous work, in particular by considering explosions that produce lower and higher ejecta kinetic energy than the canonical $1.2\times 10^{51}$\,erg. The overall approach is identical to that presented in \citet{d17_13fs} and more recently in \citet{dessart_wynn_23}, to which the reader is referred for additional details. To avoid redundancy, we thus only describe what distinguishes the present simulations from those previous studies, namely the initial conditions for the progenitor and the explosion, and the numerical setup for the \heracles\ and the \cmfgen\ simulations.

The present simulations are based on a grid of massive-star models evolved with \mesa\ \citep{mesa1,mesa2,mesa3} as part of a separate project looking at dependencies of preSN and SN properties with metallicity (the \mesa\ simulations were done in 2020 with the version 10108). The original grid includes masses 12, 15, 20, and 25\,\msun\ on the zero age main sequence (ZAMS) but the subtle differences between the preSN models (largely limited to envelope composition at the 10\,\% level for what matters in this work; see, however, \citealt{davies_dessart_19}) led us to focus on the 15\,\msun\ model only. Specifically, our 15\,\msun\ model dies with a final mass of  12.4\,\msun\ (the ``Dutch'' mass loss recipe is used with a scaling of 0.6), an effective temperature of 3980\,K, a luminosity of 96\,000\,\lsun, a surface radius $R_\ast$ of 652\,\rsun, an H-rich envelope mass of 7.85\,\msun, an He-core mass of 4.55\,\msun, and an Fe-core mass of 1.59\,\msun.

The explosion is simulated with the radiation-hydrodynamics code \voned\ \citep{livne_93,dlw10a} by injecting uniformly a fixed power within the inner 0.05\,\msun\ above the Fe core and over a duration of 0.5\,s. In this work, the total energy deposited includes the binding energy of the overlying envelope (i.e., whose absolute value is $3.7 \times 10^{50}$\,erg) and the desired asymptotic kinetic energy of 0.6, 1.2, and $1.8 \times 10^{51}$\,erg. The explosion phase is followed with \voned\ until the shock is within a few tens of \rsun\ below the $R_\ast$ of the original \mesa\ model.

\begin{figure}
\centering
\includegraphics[width=0.8\hsize]{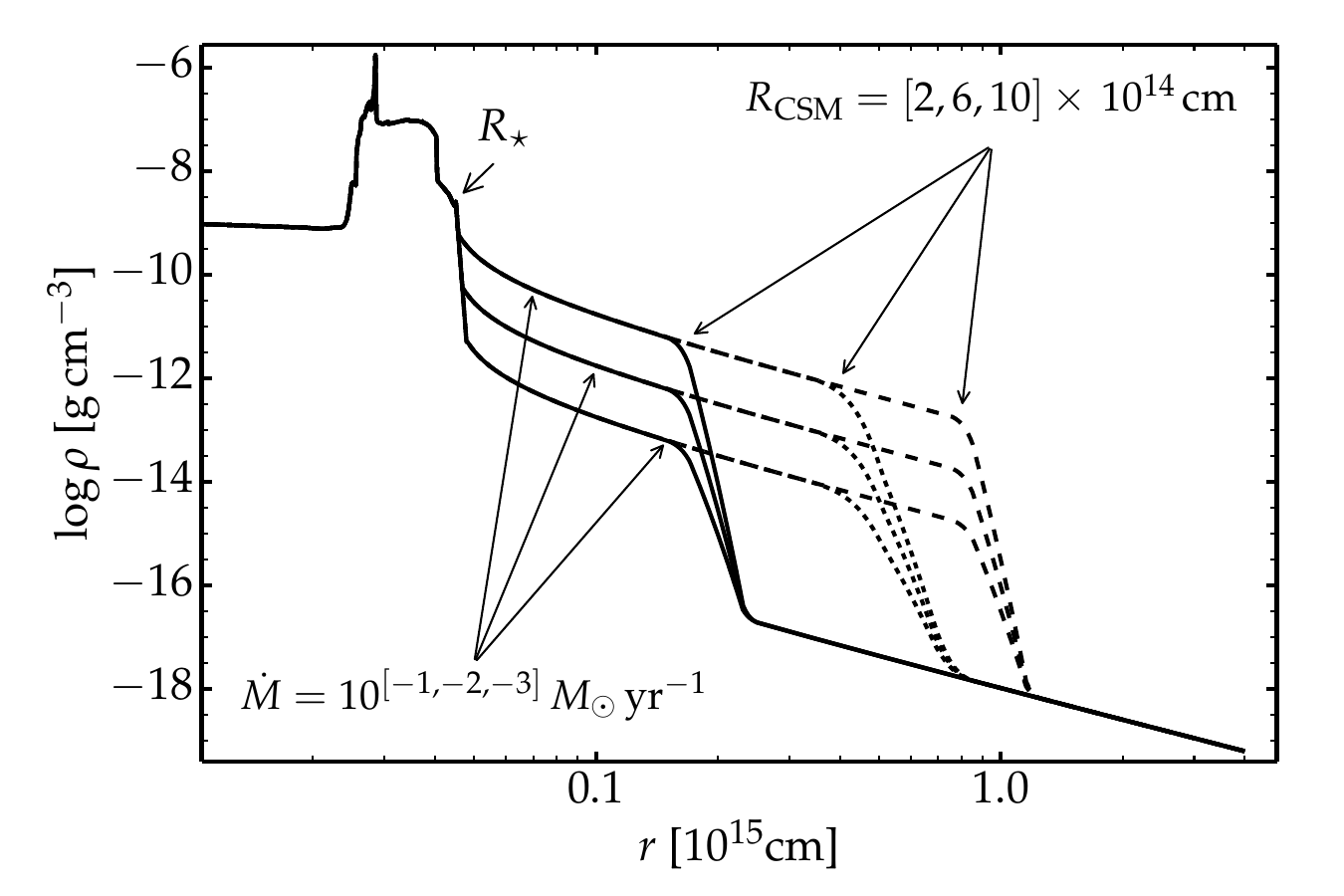}
\caption{Pre-shock breakout density profiles used as initial conditions for the radiation hydrodynamics calculations. The model is for a progenitor star of 15\,\msun\  initially and evolved at solar metallicity, which after explosion yielded an ejecta with a kinetic energy of $1.2 \times 10^{51}$\,erg. The CSM configurations are characterized by a different extent (\rcsm\ values of 2, 6, and $10 \times 10^{14}$\,cm) and density (\mdot\ values of 0.1, 0.01, and 0.001\,\msunyr). (For details, see Section~\ref{sect_setup}.)
\label{fig_init_csm}
}
\end{figure}

At that stage, the ejecta are remapped into the multi-group radiation-hydrodynamics code \heracles\ and some CSM is ``stitched'' to cover the region between $R_\ast$ and a maximum radius set at $4\times 10^{15}$\,cm. This CSM corresponds to a wind mass loss rate of 0.001, 0.01, or 0.1\,\msunyr\ out to \rcsm\ and smoothly declines within a few $10^{14}$\,cm down to a standard RSG mass loss rate of $10^{-6}$\,\msunyr. The wind velocity profile is given by $v(r) = v_0 + (v_\infty-v_0)(1-R_\ast/r)$, where $v_\infty$ is the wind terminal velocity (set to 50\,\kms\ in all cases)\footnote{Other values of $v_\infty$ and different acceleration length scales  were studied in \citet{dessart_radacc_25} but have little importance for the present work in which we focus on the information encoded in moderate- to low-resolution spectra.} and $v_0$ is the base velocity, which we adjusted so that the density profile smoothly connects with the density of the \mesa\ model at $R_\ast$. The resulting density profiles used as initial conditions for our grid of \heracles\ simulations is shown in Fig.~\ref{fig_init_csm}. Models of lower or higher $E_{\rm kin}$ differ only below $R_\ast$, though mostly in post-shock temperature -- not shown --  and also correspond to different elapsed times since core bounce (i.e., reflecting different envelope shock-crossing times).

Finally, the composition is treated in a simplified manner in the \heracles\ simulations with the use of five species only (i.e., H, He, O, Si, and Fe) mostly to capture the composition stratification and select the appropriate opacity table (twenty pre-computed tables cover from solar composition up to pure iron; for details, see \citealt{D15_2n}). This matters little here since we focus on the early time properties when only the outer ejecta composition is probed (the CSM composition is set to the surface mixture of the preSN model). In \cmfgen, we adopt a uniform composition set to the \mesa\ model at $R_\ast$. Specifically, we use $X_{\rm H}=$\,0.698, $X_{\rm He}=$\,0.287, $X_{\rm C}=$\,0.00152, $X_{\rm N}=$\,0.00228, and $X_{\rm O}=$\,0.00593, with all other abundances set to solar. Variations in metallicity have little impact for the early-time spectra discussed here because, for the corresponding high gas temperatures, iron-group-elements contribute mostly opacity in the UV and far-UV. The optical spectra thus contain only lines of H, He, C, N, and O plus continuum and essentially nothing else.

In our nomenclature, model ekin1p8\_mdot0p001\_rcsm2e14 corresponds to an ejecta with a kinetic energy of $1.8 \times 10^{51}$\,erg combined with a CSM consisting of a wind with a mass loss rate of 0.001\,\msunyr\ and extending out to $2 \times 10^{14}$\,cm. A summary of the whole model grid is given in Table~\ref{tab_init_csm}.

\begin{table}
\caption{Summary of CSM properties adopted as initial conditions for our model grid.}
\label{tab_init_csm}
\begin{center}
\begin{tabular}{cccc}
\hline
     \mdot\ & \rcsm\  &  $M_{\rm CSM}$ &  $R_{\rm phot}$  \\
$[M_\odot$\,yr$^{-1}]$ &  [10$^{14}$\,cm]     &   [\msun]   &     [10$^{14}$\,cm] \\
\hline
   0.001   &  2   &  8.62(-3)   &    1.39    \\
   0.001   &  6   &  1.07(-2)   &    2.56    \\
   0.001   & 10   &  1.35(-2)   &    3.42    \\
\hline
   0.01   &  2   &  2.40(-2)   &    1.68    \\
   0.01   &  6   &  4.43(-2)   &    4.14    \\
   0.01   & 10   &  7.19(-2)   &    7.32    \\
\hline
   0.1     &  2   &  1.75(-1)   &    1.80    \\
   0.1     &  6   &  3.72(-1)   &    4.70    \\
   0.1     & 10   &  6.50(-1)   &    8.48    \\
\hline
\end{tabular}
\end{center}
Notes: The columns give the correspondence between the CSM density (i.e., \mdot) and the extent (\rcsm) on the one hand, and the CSM mass and the radius of the electron-scattering photosphere initially on the other hand. For the latter, an ionized H-rich CSM is assumed (i.e., $\kappa=$\,0.34\,cm$^2$\,g$^{-1}$). All simulations presented here are based on a ZAMS star of 15\,\msun\ evolved at solar metallicity. (See Section~\ref{sect_setup} for discussion.)
\end{table}

\begin{figure*}
\centering
\includegraphics[width=0.8\hsize]{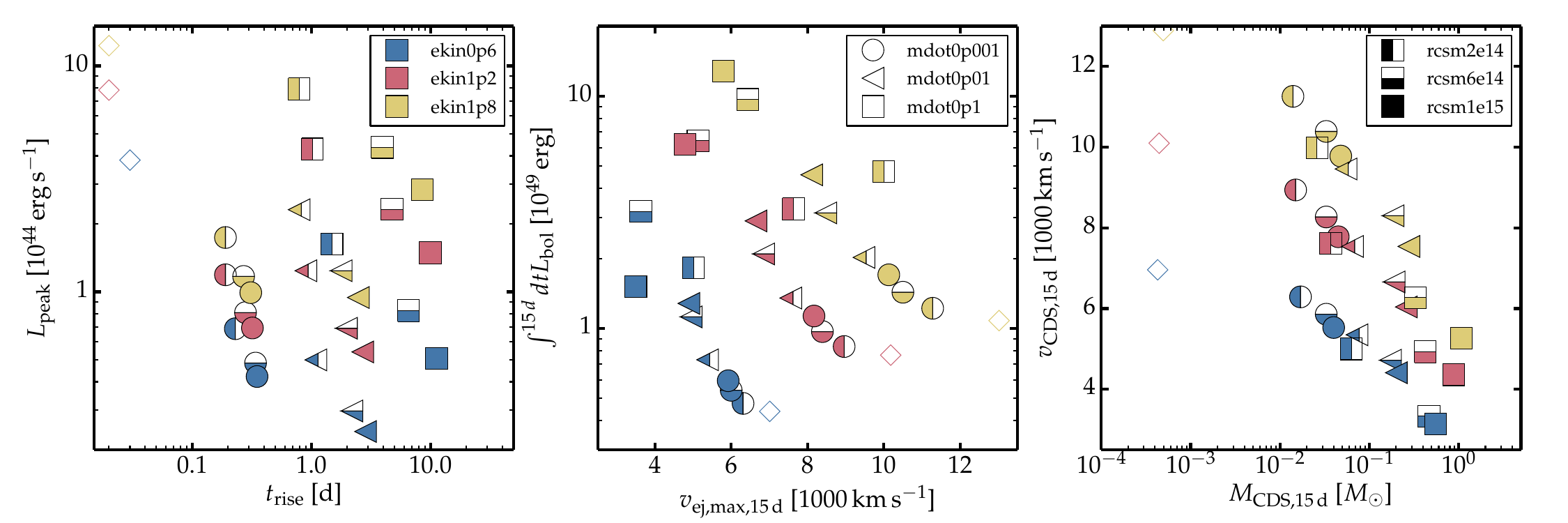}
\caption{Illustration of results from the radiation-hydrodynamics calculations for our sample of 27 ejecta/CSM configurations. Left: Peak luminosity versus rise time to peak. Middle: Time-integrated bolometric luminosity versus maximum ejecta velocity, both evaluated at 15\,d. Right: CDS velocity versus CDS mass at 15\,d. Colors differentiate different model kinetic energies, symbols the mass-loss rate, and the filling style the CSM extent. These indications, which apply to all panels, are given separately, one panel at a time for better visibility. For comparison, we also show the results for the three No-CSM cases (empty diamonds). The interaction phase is not over in models with large CSM mass and extent so the values are not yet converged at 15\,d in those cases.
\label{fig_rhd_res}
}
\end{figure*}

\begin{figure*}
\centering
\includegraphics[width=0.7\hsize]{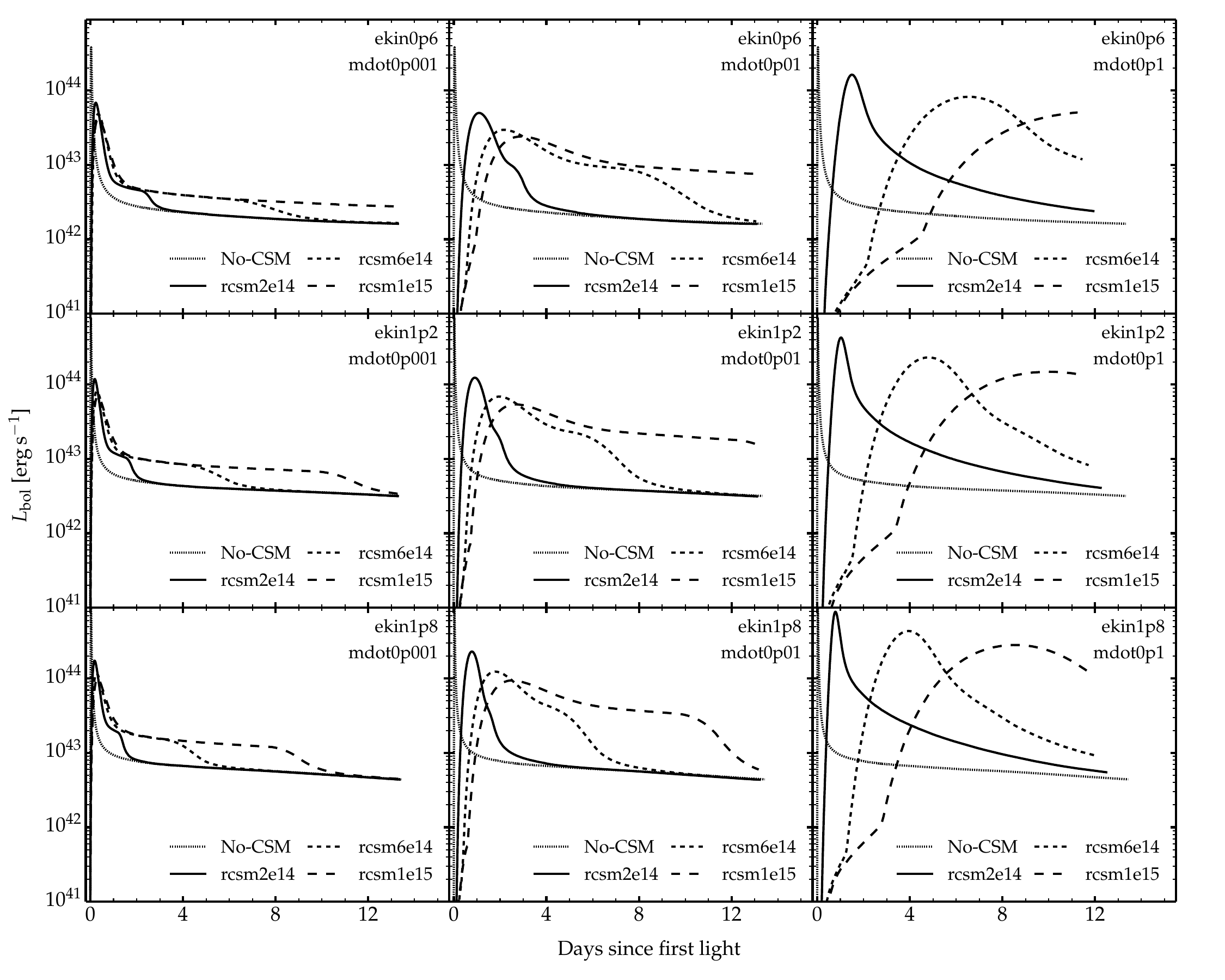}
\caption{Bolometric light curves calculated with the multi-group radiation-hydrodynamics code \heracles\ and tiled according to the model ejecta kinetic energy (one value per row), CSM density (one value of \mdot\ per column), and CSM extent (three values per panel, with the no-CSM counterpart shown as a dotted line). The x-axis origin is chosen to be when each model first brightens to a luminosity of 10$^{40}$\,\ergs, as recorded at the outer grid boundary.
\label{fig_lbol_heracles}
}
\end{figure*}

\begin{table*}
\caption{Summary of light-curve and ejecta properties from the \heracles\ simulations for our model grid. For each group of models, we also include the counterpart without CSM (i.e., \mdot\,$=$\,10$^{-6}$\,$M_\odot$\,yr$^{-1}$). \label{tab_rhd_res} }
\begin{center}
\begin{tabular}{cccllcccccc}
\hline
$E_{\rm kin}$  &  \mdot\  &   \rcsm\       & $t_{\rm rise}$  & $L_{\rm peak}$ & $\int^{15\,d}\,dt\,L_{\rm bol}$
& $V_{\rm ej, max, 15\,d}$ &    $V_{\rm CDS, 15\,d}$ & $M_{\rm CDS, 15\,d}$  &  $\tau_{\rm CDS, 15\,d}$ & $\tau_{\rm CSM, 15\,d}$  \\
\hline
  [foe] & [$M_\odot$\,yr$^{-1}$]   &    [10$^{14}$\,cm]  &  [d] &      [\ergs] & [erg] & [\mms] & [\mms]  &    [\msun] &    &   \\
\hline
   0.6  &   1(-6)   &$\cdots$&   0.03   &   3.83(44)  &    4.39(48)  &      7.01 &     6.96 &   4.28(-4) &    1.92(-2)&     Thin \\
\hline
   0.6  &   0.001   &   2    &   0.23   &   6.87(43)  &    4.75(48)  &      6.31 &     6.29 &      0.017 &       0.83 &     Thin   \\
   0.6  &   0.001   &   6    &   0.34   &   4.84(43)  &    5.41(48)  &      6.00 &     5.86 &      0.033 &       1.51 &     Thin   \\
   0.6  &   0.001   &   10   &   0.35   &   4.22(43)  &    5.95(48)  &      5.92 &     5.53 &      0.040 &       2.41 &     Thin    \\
\hline
   0.6  &    0.01   &   2    &   1.09   &   5.00(43)  &    7.32(48)  &       5.42 &    5.26 &      0.058 &         5.01 &    Thin   \\
   0.6  &    0.01   &   6    &   2.19   &   2.97(43)  &    1.12(49)  &       4.95 &    4.72 &      0.173 &        16.31 &    Thin   \\
   0.6  &    0.01   &   10   &   2.85   &   2.41(43)  &    1.28(49)  &       4.89 &    4.41 &      0.201 &        20.78 &    0.99  \\
\hline
   0.6  &     0.1   &   2    &   1.49   &   1.63(44)  &    1.82(49)  &       5.01 &    5.00 &      0.063 &         5.96 &    Thin   \\
   0.6  &     0.1   &   6    &   6.55   &   8.26(43)  &    3.19(49)  &       3.63 &    3.33 &      0.467 &        82.55 &    0.53   \\
   0.6  &     0.1   &   10   &  11.25   &   5.07(43)  &    1.51(49)  &       3.50 &    3.14 &      0.551 &       107.35 &    28.5  \\
\hline
\hline
   1.2  &  1(-6)    &$\cdots$&   0.02   &   7.82(44)  &    7.67(48)  &      10.18 &    10.10&    4.44(-4)&      9.84(-3)&    Thin \\
\hline
   1.2  &   0.001   &   2    &   0.19   &   1.19(44)  &    8.35(48)  &       9.03 &    8.83 &      0.012 &         0.40 &    Thin   \\
   1.2  &   0.001   &   6    &   0.28   &   8.09(43)  &    9.68(48)  &       8.47 &    8.24 &      0.029 &         0.98 &    Thin   \\
   1.2  &   0.001   &   10   &   0.32   &   6.92(43)  &    1.13(49)  &       8.29 &    7.73 &      0.039 &         1.37 &    Thin   \\
\hline
   1.2  &    0.01   &   2    &   0.90   &   1.24(44)  &    1.35(49)  &       7.56 &    7.54 &      0.065 &         2.39 &    Thin   \\
   1.2  &    0.01   &   6    &   1.97   &   6.90(43)  &    2.09(49)  &       6.85 &    6.66 &      0.190 &         9.33 &    Thin   \\
   1.2  &    0.01   &   10   &   2.70   &   5.42(43)  &    2.90(49)  &       6.65 &    6.04 &      0.260 &        14.67 &    Thin   \\
\hline
   1.2  &     0.1   &   2    &   1.02   &   4.29(44)  &    3.27(49)  &       7.63 &    7.62 &      0.037 &         1.53 &    Thin   \\
   1.2  &     0.1   &   6    &   4.75   &   2.32(44)  &    6.42(49)  &       5.13 &    4.93 &      0.425 &        38.47 &    Thin   \\
   1.2  &     0.1   &   10   &   9.98   &   1.49(44)  &    6.21(49)  &       4.79 &    4.36 &      0.879 &        96.61 &    11.3  \\
\hline
\hline
   1.8  &  1(-6)    &$\cdots$&   0.02   &   1.23(45)  &    1.08(49)  &      13.02 &   12.89 &   4.95(-4) &      6.74(-3)&    Thin \\
\hline
   1.8  &   0.001   &   2    &   0.19   &   1.74(44)  &    1.22(49)  &      11.28 &   11.26 &      0.014 &         0.24 &    Thin  \\
   1.8  &   0.001   &   6    &   0.27   &   1.17(44)  &    1.43(49)  &      10.50 &   10.39 &      0.033 &         0.64 &    Thin  \\
   1.8  &   0.001   &   10   &   0.31   &   9.92(43)  &    1.70(49)  &      10.13 &    9.78 &      0.048 &         0.98 &    Thin   \\
\hline
   1.8  &    0.01   &   2    &   0.79   &   2.31(44)  &    2.02(49)  &       9.51 &    9.29 &      0.043 &         1.34 &    Thin   \\
   1.8  &    0.01   &   6    &   1.78   &   1.24(44)  &    3.14(49)  &       8.48 &    8.30 &      0.185 &         5.63 &    Thin   \\
   1.8  &    0.01   &   10   &   2.48   &   9.44(43)  &    4.58(49)  &       8.41 &    7.53 &      0.212 &        10.04 &    Thin   \\
\hline
   1.8  &     0.1   &   2    &   0.79   &   7.91(44)  &    4.73(49)  &      10.00 &    9.99 &      0.026 &         0.62 &    Thin   \\
   1.8  &     0.1   &   6    &   3.94   &   4.36(44)  &    9.67(49)  &       6.43 &    6.27 &      0.328 &        19.16 &    Thin   \\
   1.8  &     0.1   &   10   &   8.56   &   2.83(44)  &    1.28(50)  &       5.79 &    5.27 &      1.072 &        79.70 &    3.23  \\
\hline
\end{tabular}
\end{center}
Notes: For each model characterized by a value of $E_{\rm kin}$, \mdot, and \rcsm, the columns give the rise time to bolometric maximum (from a value of 10$^{40}$\,\ergs; $t_{\rm rise}$), the peak bolometric luminosity ($L_{\rm peak}$), followed by quantities evaluated at 15\,d including the time-integrated bolometric luminosity ($\int^{15\,d}\,dt\,L_{\rm bol}$), the maximum ejecta velocity ($V_{\rm ej, max, 15\,d}$), the CDS velocity ($V_{\rm CDS, 15\,d}$), the CDS mass ($M_{\rm CDS, 15\,d}$), the optical depth across the CDS ($\tau_{\rm CDS, 15\,d}$), and the optical depth of the unshocked CSM ($\tau_{\rm CSM, 15\,d}$).
\end{table*}

\begin{figure}[h]
\centering
\includegraphics[width=0.8\hsize]{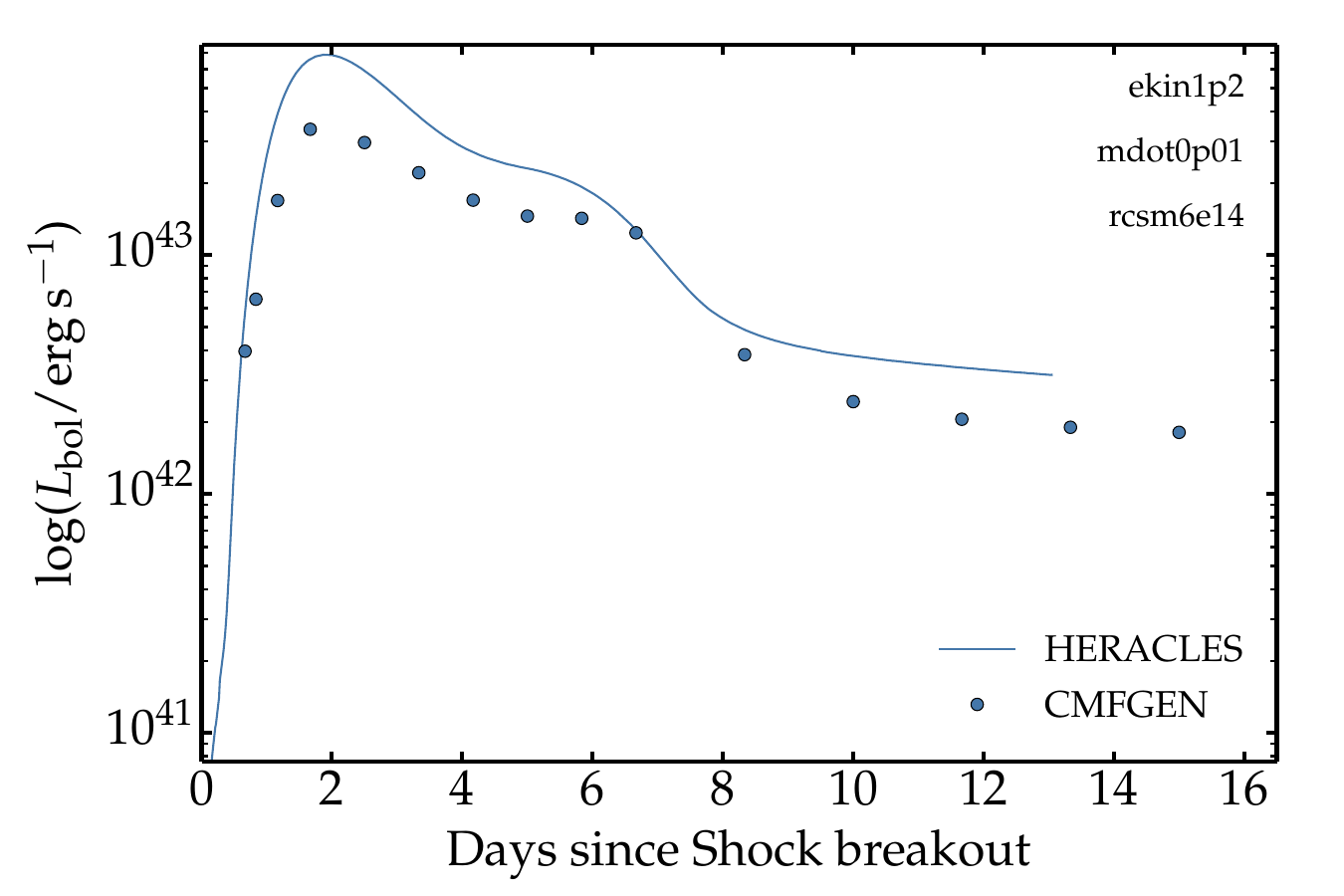}
\caption{Comparison between the \heracles\ and \cmfgen\ bolometric light curves for model ekin1p2\_mdot0p01\_rcsm6e14. The \heracles\ light curve has been shifted in time to correct for the light-travel time to the outer grid boundary.
\label{fig_comp_heracles_cmfgen}
}
\end{figure}

\section{Results from the radiation-hydrodynamics simulations with \heracles.}
\label{sect_rhd}

Figure~\ref{fig_rhd_res} summarizes the results from the \heracles\ simulations, specifically showing the variations in peak bolometric luminosity relative to the rise time to peak, the time-integrated bolometric luminosity up to 15\,d relative to the maximum ejecta velocity at 15\,d, and the CDS velocity versus the CDS mass at 15\,d (see also Table~\ref{tab_rhd_res}). Similar properties were shown in \citet{dessart_wynn_23} for a model set limited to just a few CSM (or associated mass loss rate) configurations.

The large range in ejecta kinetic energy of our model sample yields a broad range of outcomes (e.g., broader than in the simulations of \citealt{dessart_wynn_23}) because the ejecta kinetic energy is the energy source for the system (left panel of Fig.~\ref{fig_rhd_res}). The rise time to peak covers from a small fraction of a day (cases of small \mdot, small \rcsm, high \ekin) up to about 12\,d, and reflects both the variation in shock-crossing time through the CSM and the diffusion time. The model ekin0p6\_mdot0p1\_rcsm1e15 is the one with the longest rise time not only because the SN shock is slow moving to start with but it is also strongly decelerated by a dense and extended CSM. The peak luminosity is typically of the order of 10$^{44}$\,\ergs. All else being kept the same, a greater \rcsm\ or a greater \mdot\ have a qualitatively similar impact. The enhanced diffusion time tends to decrease $L_{\rm peak}$ (the stored radiation energy is released over a longer duration) but the greater extraction of ejecta kinetic energy tends to increase $L_{\rm peak}$. What results in practice is case dependent, but in our simulations, all configurations with CSM have a lower peak luminosity than in the No-CSM case (which is of order 10$^{45}$\,\ergs\ -- the rise time in the No-CSM case is only about one hour and the peak is very narrow). For the models mdot0p1 and rcsm1e15, the rise time is $\sim$\,10\,d and thus the interaction is still ongoing at 15\,d. Such models belong to Type IIn SNe in which interaction may continue for weeks and produce a superluminous transient.

Extraction of ejecta kinetic energy implies a deceleration of the outer ejecta and a reduction of the maximum ejecta velocity (middle panel of Fig.~\ref{fig_rhd_res}). It covers from $\sim$\,6300 to 3500\,\kms\ in models with ekin0p6, from $\sim$\,9000 to 4800\,\kms\ in models with ekin1p2, and from 11\,500 to 5800\,\kms\ in models with ekin1p8. All else being kept fixed, a greater CSM extent or mass causes a greater reduction in $V_{\rm ej,max,15\,d}$. Commensurate with this reduction of $V_{\rm ej,max,15\,d}$ (or outer ejecta kinetic energy) is the time-integrated bolometric luminosity. In the ekin0p6 model set, the value relative to the No-CSM case (i.e., $4.39 \times 10^{48}$\,erg) spans from $4.75 \times 10^{48}$\,erg up to $3.19 \times 10^{49}$\,erg. In the ekin1p2 model set, the value relative to the No-CSM case (i.e., $7.67 \times 10^{48}$\,erg) spans from $8.35 \times 10^{48}$\,erg up to $6.42 \times 10^{49}$\,erg. In the ekin1p8 model set, the value relative to the No-CSM case (i.e., $1.08 \times 10^{49}$\,erg) spans from $1.22 \times 10^{49}$\,erg up to $1.28 \times 10^{50}$\,erg. In models mdot0p1 and rcsm1e15, the time-integrated bolometric luminosity is still growing significantly at 15\,d due to the persisting interaction (the value at 15\,d may thus be close to or below that of the corresponding model with rcsm6e14).

The CDS mass  at 15\,d spans a wide range of values, from $\sim$\,0.01\,\msun\ in the mdot0p001 cases (barely above the No-CSM case) up to $\sim$\,1\,\msun\ in the mdot0p1 cases, wherein the interaction is still ongoing at 15\,d (right panel of Fig.~\ref{fig_rhd_res}). The greater the value of $M_{\rm CDS,15\,d}$, the lower the value of the CDS velocity at 15\,d, which spans 6300 down to 3100\,\kms\ in the model set ekin0p6, 8900 down to 4300\,\kms\ in the model set ekin1p2, and 11\,300 down to 5300\,\kms\ in the model set ekin1p8.

Bolometric light curves for the full set of simulations with \heracles\ are shown in Fig.~\ref{fig_lbol_heracles}. The morphology is similar in all cases with a peak broadened over a diffusion time (i.e., broader for greater CSM mass or extent), and a luminosity that settles onto a ledge as the shock progresses through the outer regions of the dense CSM, before marking an abrupt drop as the shock exits this dense CSM around \rcsm. An interesting feature particularly visible in the models mdot0p1 and rcsm1e15 is the break in light-curve slope that occurs after a few days, hence on the rise to peak, at a power of about 10$^{42}$\,\ergs. This change in slope arises from the arrival (at the outer grid radius where the luminosity is recorded) of photons with a wavelength longer than the Balmer edge. It also coincides with the time when the CSM is ionized throughout. Thus, prior to this kink, the luminosity of the SN arises exclusively from low energy (non-ionizing) photons beyond 3670\,\AA, and the kink signals a sudden change not just in luminosity, but also in color (i.e., bluer) and ionization (i.e., ionized species start to appear). This transition may be at the origin of a similar light curve kink and color change observed in SN\,2024ggi \citep{shrestha_24ggi_24,zhang_24ggi_24,wynn_24ggi_24}.

Although the dense CSM has been swept-up in most simulations at 15\,d (tag ``Thin'' in the column $\tau_{\rm CSM,15\,d}$), which corresponds to the end of the ``IIn phase'' \citep{d17_13fs,dessart_wynn_23,wynn_pap2_24}, there remains some residual, optically-thick, unshocked CSM in models mdot0p1 and rcsm1e15. In all other cases, the system is in the ``CDS phase'', meaning that the photosphere (and the spectrum formation region) lies in the CDS \citep{D16_2n,dessart_review_26,zheng_23ixf_25}. This is not surprising since even in RSG star explosions occurring in a vacuum, the photosphere lies no more than a few 0.01\,\msun\ below the progenitor surface within the first 15\,d after explosion \citep{bersten_11_2p,DH11_2p}.


\begin{figure*}[h]
\centering
\includegraphics[width=0.7\hsize]{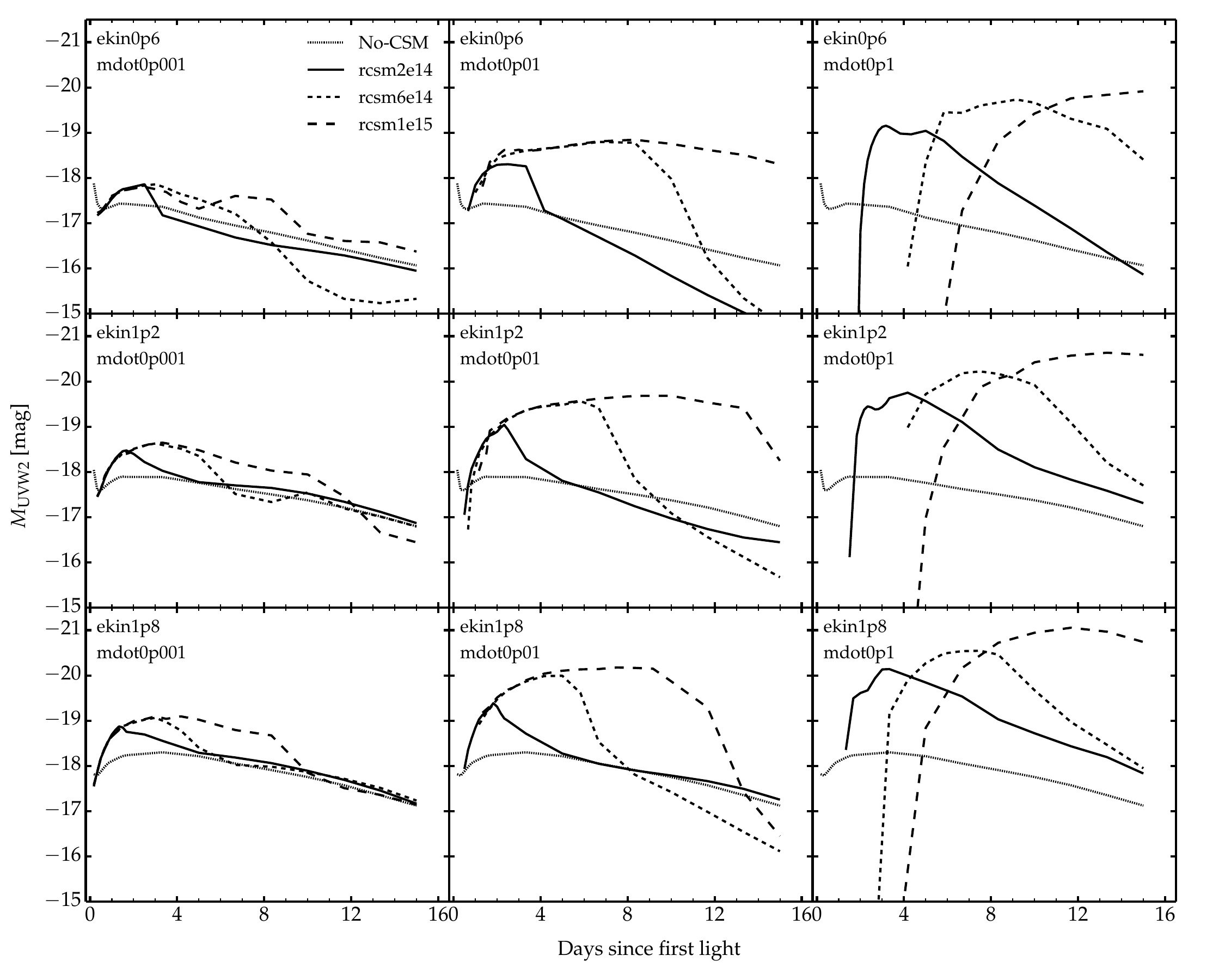}
\caption{Same as Fig.~\ref{fig_lbol_heracles} but now showing the $UVW2$-band light curves computed with \cmfgen\ for our model set.
\label{fig_lc_uvw2_cmfgen}
}
\end{figure*}

\section{Multi-band light curves computed with \cmfgen.}
\label{sect_lc_cmfgen}

  In this section, we discuss the results from the radiative-transfer calculations performed with \cmfgen\ on selected snapshots of the 27 ejecta/CSM configurations modeled with \heracles, with a focus on the photometric properties. As indicated in Section~\ref{sect_setup} and in \citet{D15_2n,d17_13fs}, the Sobolev, nonmonotonic-velocity solver in the NLTE code \cmfgen\ is used to post-process the simulations from the radiation-hydrodynamics code \heracles. Because the temperature structure is strongly influenced by the dynamics (e.g., the location of the shock, its propagation, the optical depth of the surrounding material etc), we import that temperature in \cmfgen\ and hold it fixed. But unlike \heracles, which assumes LTE for the gas (e.g., its ionization level is fully known from the composition, the temperature, and the electron density), \cmfgen\ does not typically find that cooling and heating rates are equal with this adopted temperature. This inconsistency is the main reason behind the systematic offset of up to about 80\,\% of the luminosity obtained with \cmfgen\ compared to that of \heracles\ for a given snapshot. Figure~\ref{fig_comp_heracles_cmfgen} illustrates this offset between the two codes for model ekin1p2\_mdot0p01\_rcsm6e14. Thus, the magnitudes obtained with \cmfgen\ are likely underestimated by 0.5--1.0\,mag.

Figure~\ref{fig_lc_uvw2_cmfgen} illustrates the $UVW2$-band light curves for the full model grid and using a similar tiling as for the bolometric light curves obtained with \heracles\ and shown in Fig.~\ref{fig_lbol_heracles}. Compared to the No-CSM counterparts, models with interaction exhibit a significant boost to the $UVW2$ brightness and by extension to the UV luminosity, and the more so for denser and more extended CSM. Models with modest CSM mass or extent peak at about $-18.5$\,mag whereas models with high CSM mass or extent peak at about $-20.5$\,mag. This boost persists for a fews days in rcsm2e14 models but for 1--2 weeks in rcsm1e15 models. This overall pattern is however mitigated by at least two factors. First, the diffusion of SN radiation through the optical-thick CSM can delay the initial (i.e., the actual breakout) brightening by several days compared to the No-CSM counterpart in which the breakout signal peaks within about an hour of shock emergence. The subsequent signal is then not just composed of the broadened, diffuse breakout burst but also augmented by the continuous supply of power from the interaction of the ejecta with this CSM, as well as the release of shock-deposited energy from the outermost ejecta layers. Secondly, the formation of the dense shell can lead to enhanced blanketing at later times, eventually causing a drop of the $UVW2$ brightness. This feature of the model is sensitive to the adopted CSM density profile -- here we impose a strong and prompt reduction of the CSM density beyond \rcsm. A more smoothly declining CSM density profile would produce a more smoothly decline $UVW2$-band brightness, as observed \citep{wynn_pap2_24}.

Figure~\ref{fig_lc_uvw2_cmfgen} indicates that the $UVW2$ brightness is a sensitive probe of ejecta interaction with CSM, with a magnitude modulation that typically survives on the shock-crossing time through the dense CSM and scales by $\lesssim$\,1\,mag, $\sim$\,1.5\,mag, and even $\gtrsim$\,2\,mag as the wind mass loss rate is increased from 0.001 to 0.01 and 0.1\,\msunyr. The results for the $V$-band light curves (shown in the appendix in Fig.~\ref{fig_lc_v_cmfgen}) reveal a weaker impact of CSM, likely because the optical range probes the Rayleigh-Jeans tail of the spectral energy distribution (i.e., the hot emitting plasma radiates mostly in the UV). Nonetheless, one important impact of the CSM is to shorten the rise time in the optical. The short $V$-band rise-time of even standard Type II SNe suggests that some CSM, even if compact, is likely present in most or all exploding RSG stars \citep{gonzalez_gaitan_2p_15,morozova_2l_2p_17,hinds_csm_25}.


\section{Results for the spectroscopic properties}
\label{sect_spec}

In this section, we discuss the results from the radiative-transfer calculations performed with \cmfgen\ but with a  focus on the spectral evolution for the full model grid. The main focus is on the IIn-phase which varies in duration depending on the ejecta kinetic energy, CSM mass and extent of the configuration.

\begin{figure*}
\centering
\includegraphics[width=0.7\hsize]{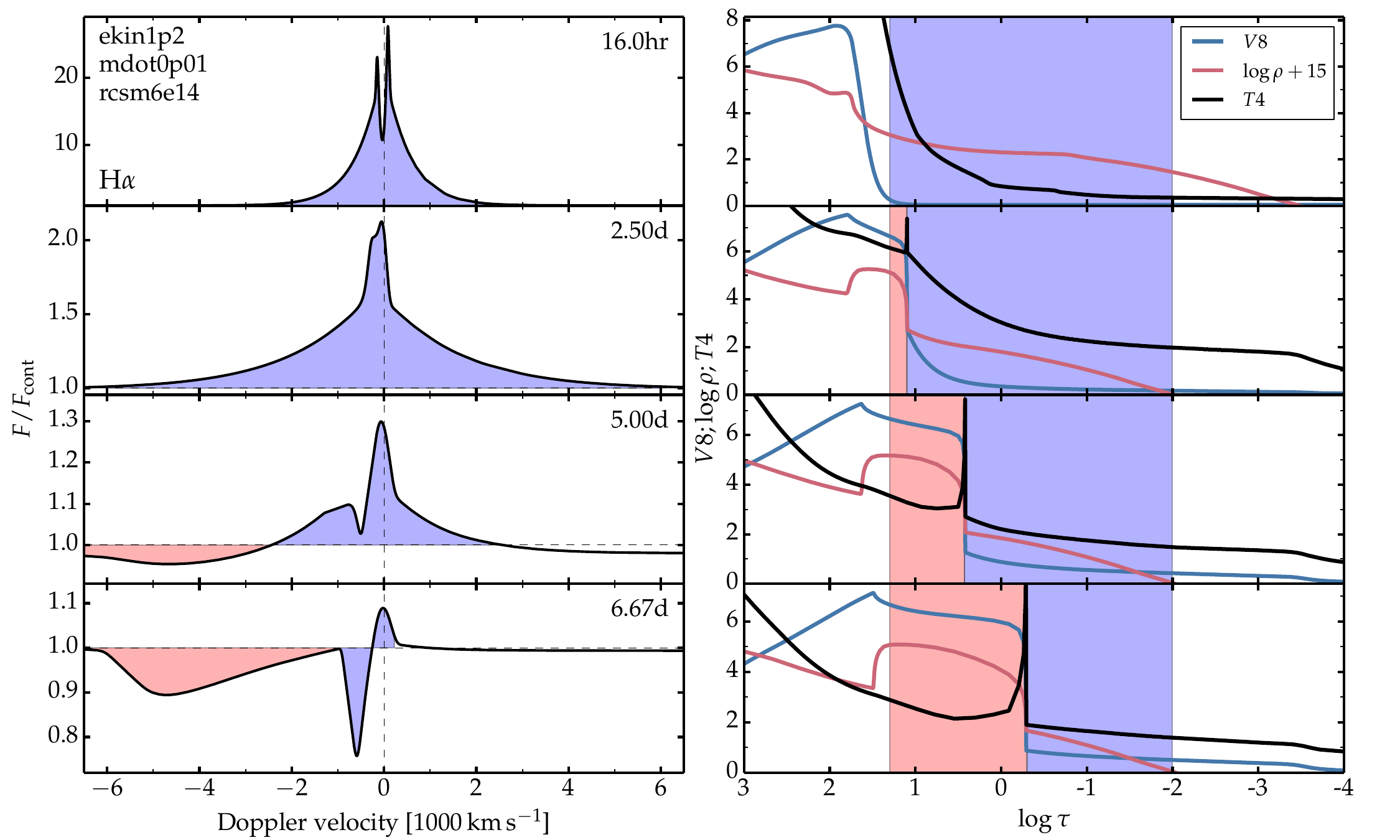}
\caption{Illustration of the correspondence between the H$\alpha$ evolution (left column; only H\one\ lines are included in the displayed range) and the dynamical properties in the spectrum formation region (right column) for model ekin1p2\_mdot0p01\_rcsm6e14. The shading differentiates emission from the unshocked CSM (blue) and the shocked material (red), which corresponds here to the dense shell. The spectrum forms preferentially over regions where $\log \tau$ is between $-1$ and $1$, except for the narrow line core arising from regions at $\log \tau$ of about $-2$. Here, $V8$ refers to the velocity in units of $10^8$\,cm\,s$^{-1}$ and $T4$ is the gas (or electron) temperature in units of $10^4$\,K ($\log \rho$ is the base-10 logarithm of the density, shown with an additive constant for better visibility).
\label{fig_halpha_ekin1p2_mdot0p01_rcsm6e14}
}
\end{figure*}

\subsection{H$\alpha$ as diagnostic of the dynamics}
\label{sect_ha}

In all the simulations performed, the only line present and strong at all epochs (either in emission or absorption) is H$\alpha$ (and to a lesser extent H$\beta$). Compared to other species, there is only one higher ionization state (the same applies to He\two) and consequently H\one\ lines are present irrespective of the ionization level or temperature of the gas. This property also implies that H\one\ lines may form over a large range of depths even in the presence of an ionization stratification. These properties thus make H$\alpha$ an ideal diagnostic of the dynamics of the interaction of SN ejecta with CSM. Here, we focus primarily the discussion on the ``broad'' component of H$\alpha$, as would be observed in low-resolution spectra -- a discussion of the narrow component (observable in high-resolution spectra) is given in \citet{dessart_radacc_25}.

Figure~\ref{fig_halpha_ekin1p2_mdot0p01_rcsm6e14} illustrates in the left column the H$\alpha$ evolution in model ekin1p2\_mdot0p01\_rcsm6e14 over the first week as the SN evolves through the IIn phase (first two epochs) into the CDS phase (last two epochs), together with the corresponding ejecta/CSM structure (velocity, density, and temperature) from the \heracles\ simulation at the corresponding epoch in the right column. Using a colored shading, we indicate the spatial regions (right) at the origin of the emission or absorption regions in H$\alpha$ (left). For the first epoch, H$\alpha$ exhibits a symmetric, electon-scattering broadened emission profile with a narrow core. The flux in the extended line wings arise from slow-moving ($<$\,50\,\kms), relatively cool ($\sim$\,10\,000\,K), unshocked CSM wherein the radiation from the shock is absorbed and remitted, as well as scattered by free electrons. The narrow line core arises from an external region at low electron-scattering optical depth where photons suffer essentially no frequency redistribution by free electrons. Thus at this first epoch, the entire line emission arises from unshocked CSM. Radiative acceleration of the unshocked CSM also leads to the disappearance of the SN shock at this early time.

At the second epoch of 2.5\,d shown in the second row of Fig.~\ref{fig_halpha_ekin1p2_mdot0p01_rcsm6e14}, the configuration is essentially unchanged apart from a clear radiative acceleration of the unshocked CSM, with velocities as high as $\sim$\,2000\,\kms\ at $\tau$\,$\sim$\,10 (just exterior to the SN shock, which is now visible), 220\,\kms\ at $\tau$\,$\sim$\,0.1, and 170\,\kms\ at $\tau$\,$\sim$\,0.01 where the narrow line core (though now broadened) forms (see discussion in \citealt{dessart_radacc_25}).

At the third epoch of 5.0\,d shown in the third row of Fig.~\ref{fig_halpha_ekin1p2_mdot0p01_rcsm6e14}, the configuration is now qualitatively changed because the profile combines emission from unshocked CSM (whose volume decreased due to the shock expansion) and absorption from the fast moving dense shell located at an optical depth of a few (red shading). This well-formed dense shell is easily identifiable as the density bump with a representative velocity of 6000\,\kms\ (right column of Fig.~\ref{fig_halpha_ekin1p2_mdot0p01_rcsm6e14}).

At the last epoch of 6.67\,d shown in the bottom row of Fig.~\ref{fig_halpha_ekin1p2_mdot0p01_rcsm6e14}, the unshocked CSM is now entirely optically thin (in electron scattering) and contributes only narrow emission at line center. The SN radiation has, however, accelerated that material around $\tau=$\,0.01 up to 500\,\kms, so that this narrow line core is much broader than in the first epoch (top row). The bulk of the spectrum (around the photosphere) is now forming in the dense shell (red shading), causing an extended absorption out to $-6000$\,\kms\ on the blue side of the H$\alpha$ profile.

Our set of 27 models exhibit evolving dynamical configurations and thus evolving H$\alpha$ profile morphologies but the basic features discussed above apply in all cases. In the appendix, we show in Fig.~\ref{fig_halpha_ekin1p2_mdot0p1_rcsm2e14} the corresponding results for model ekin1p2\_mdot0p1\_rcsm2e14 in which the high CSM density but smaller CSM extent lead to a much stronger radiative acceleration of the unshocked material (see discussion in \citealt{dessart_radacc_25}). This acceleration is so strong that the SN shock essentially disappears during the first few days after the shock crosses the progenitor $R_\ast$. It leads to H$\alpha$ emission around 4000\,\kms, essentially all blue-shifted relative to line center because of optical-depth effects.

The H$\alpha$ evolution observed in SN\,1998S \citep{leonard_98S_00,fassia_98S_01} shows a qualitatively similar behavior to that shown above for model ekin1p2\_mdot0p01\_rcsm6e14, with a somewhat longer IIn phase of about a week and a prolonged CDS phase lasting until about three weeks. The high CDS mass in SN\,1998S facilitates the identification of the blue-shifted absorption from the CDS, which is observed simultaneously at day 25 (see Fig.~5 of \citealt{leonard_98S_00}) with the narrow emission and absorption from distant unshocked CSM. More frequently, this hybrid profile morphology is less obvious (see, e.g., for SNe\,2023ixf and 2024ggi; \citealt{bostroem_23ixf_23,jacobson_galan_23ixf_23,singh_23ixf_24,wynn_24ggi_24,zhang_24ggi_24,zheng_23ixf_25}), which may arise from a combination of a lower CDS mass and departures from spherical symmetry, as well as insufficient data quality (i.e., signal-to-noise ratio and spectral resolution).

\begin{sidewaystable*}
\caption{Times over which specific optical, electron-scattering broadened emission lines are present in the model spectra computed for our 27 ejecta/CSM configurations characterized by different values of $E_{\rm kin}$, \mdot, and \rcsm.}
\label{tab_spec_evol_sum}
\begin{center}
\begin{tabular}{|c@{\hspace{1mm}}|c@{\hspace{1mm}}|c@{\hspace{1mm}}|c@{\hspace{1mm}}|c@{\hspace{1mm}}|c@{\hspace{1mm}}|c@{\hspace{1mm}}|c@{\hspace{1mm}}|c@{\hspace{1mm}}|c@{\hspace{1mm}}|c@{\hspace{1mm}}|c@{\hspace{1mm}}|c@{\hspace{1mm}}|}
\hline
\rotatebox[origin=c]{90}{$E_{\rm kin}$ [foe]} &  \rotatebox[origin=c]{90}{\mdot\ [$M_\odot$\,yr$^{-1}$]} &   \rotatebox[origin=c]{90}{\rcsm\ [10$^{14}$\,cm]}   &
\rotatebox[origin=c]{90}{ IIn phase }  &
\rotatebox[origin=c]{90}{ He\,{\sc i}\,5875.66\,\AA}  &      
\rotatebox[origin=c]{90}{ He\,{\sc ii}\,4685.70\,\AA}  &     
\rotatebox[origin=c]{90}{ C\,{\sc iii}\,5695.92\,\AA}  &     
\rotatebox[origin=c]{90}{ C\,{\sc iv}\,5804.86\,\AA}  &      
\rotatebox[origin=c]{90}{``N\,{\sc iii}--C\,{\sc iii}''}  &     
\rotatebox[origin=c]{90}{ N\,{\sc iv}\,7109.35\,\AA}  &      
\rotatebox[origin=c]{90}{ N\,{\sc v}\,4612.72\,\AA}  &       
\rotatebox[origin=c]{90}{ O\,{\sc v}\,5597\,\AA}  &          
\rotatebox[origin=c]{90}{ O\,{\sc vi}\,3820\,\AA}  \\        
\hline
   0.6  &   0.001   &   2    & $\le$\,2.1 &$\cdots$ &[0.3-1.7]&$\cdots$ &[0.8-1.3]&[1.0-1.7]&[0.8-1.2]&[0.6-0.7]&0.3/0.7  & 0.5 \\
   0.6  &   0.001   &   6    & $\le$\,3.8 &$\cdots$ &[0.3-2.3]&[1.5-2.2]&[0.7-1.2]&[1.0-2.0]&[0.8-1.0]& 0.6     &0.6      & 0.5 \\
   0.6  &   0.001   &   10   & $\le$\,5.2 &$\cdots$ &[0.5-2.0]&$\cdots$ &[0.5-1.0]&[1.0-2.0]&[0.8-1.0]&[0.5-0.7]&[0.5-0.7]& 0.5  \\
\hline
   0.6  &    0.01   &   2    & $\le$\,3.2 &$\cdots$        &[0.6-3.0]&$\cdots$ &0.6/[2.0-2.5]&0.6/2.5       &0.6/[2.0-2.5]&[1.0-2.0]&[1.0-1.7]&[1.0-1.4] \\
   0.6  &    0.01   &   6    & $\le$\,8.8 &[4.0-6.7]       &[1.0-6.7]&[4.0-6.7]&[1.3-3.5]    &1.0/[3.3-5.0] &[1.3-3.3]    &2.0      &$\cdots$ &$\cdots$  \\
   0.6  &    0.01   &   10   & $>$\,15.0  &1.3/[4.0-12.0]  &[1.3-6.0]&[5.0-7.0]&[1.7-3.3]    &[1.3-6.0]     &$\cdots$     &$\cdots$ &$\cdots$ &$\cdots$  \\
\hline
   0.6  &     0.1   &   2    & $\le$\,4.7 &2.0                  &[2.0-4.5]  &2.0        &2.2/[3.8-4.3] &[2.0-2.3]    &[2.0-2.4]/[3.8-4.3] &[2.3-3.8] &[2.3-3.5]/3.3 &[2.6-3.3]   \\
   0.6  &     0.1   &   6    & $\le$\,14.9&[4.2-5.0]/[11.5-13.3]&[5.0-11.5] &5.0/11.5   &[5.8-10.0]    &[5.0-11.5]   &[5.8-10.0]          &5.8/7.5   &[7.5-9.0]     &$\cdots$    \\
   0.6  &     0.1   &   10   & $>$\,15.0  &[6.0-10.0]           &[10.0-15.0]&[10.0-15.0]&[11.6-15.0]   &10.0-15.0]   & $\cdots$           &$\cdots$  &$\cdots$      &$\cdots$   \\
\hline
\hline
   1.2  &   0.001   &   2    & $\le$\,1.4 &$\cdots$        &[0.3-1.4] &$\cdots$ &[0.7-1.3]     &$\cdots$     &[0.8-1.2]    &[0.5-0.7] &[0.5-0.7] & 0.5   \\
   1.2  &   0.001   &   6    & $\le$\,2.7 &$\cdots$        &[0.3-2.7] &[1.6-2.3]&[0.8-1.2]     &[1.0-2.0]    &[0.7-1.2]    &[0.5-0.7] &[0.5-0.7] & 0.5   \\
   1.2  &   0.001   &   10   & $\le$\,3.7 &$\cdots$        &[0.3-2.7] &$\cdots$ &[0.8-1.2]     &[1.0-2.5]    &[0.8-1.0]    &[0.5-0.8] &[0.5-0.8] & 0.5   \\
\hline
   1.2  &    0.01   &   2    & $\le$\,2.3 &$\cdots$            &[0.5-2.2] &$\cdots$  &[0.5-1.7]/[1.5-2.2] &0.5/[2.0-2.2] &0.7/[1.8-2.2] &[0.7-1.7] &[0.8-1.7] & [1.0-1.4]   \\
   1.2  &    0.01   &   6    & $\le$\,6.3 &[0.7-0.8]/[5.0-6.0] &[0.7-5.5] &[5.0-5.5] &[1.2-3.3]           &0.8/[3.3-5.0] &[1.2-3.3]     &1.7       & $\cdots$ & $\cdots$  \\
   1.2  &    0.01   &   10   & $\le$\,11.7&1.2/[5.0-10.0]      &[1.2-6.7] &[5.0-7.0] &[1.3-3.3]           &[3.3-5.0]     &[1.7-3.3]     &$\cdots$  & $\cdots$ & $\cdots$  \\
\hline
   1.2  &     0.1   &   2    & $\le$\,3.2 &1.5             &[1.8-3.0]  &$\cdots$ &[3.0-3.3]     &1.5           &[3.2-3.3]    &1.8/[2.6-3.0] &1.8/[2.6-3.0] & [1.8-2.7]   \\
   1.2  &     0.1   &   6    & $\le$\,10.9&4.2/10.0        &[4.2-10.0] &4.2/10.0 &[5.0-8.3]     &4.2/[8.3-9.2] &5.0/[7.5-8.3]&[5.0-6.7]     &[5.0-7.5]     & 5.8  \\
   1.2  &     0.1   &   10   & $>$\,15.0  &[5.0-6.7]       &[6.7-15.0] &6.7      &[7.5-13.3]    &[6.7-15.0]    &[8.3-13.3]   &$\cdots$      &$\cdots$      & $\cdots$   \\
\hline
\hline
   1.8  &   0.001   &   2    & $\le$\,1.1 &$\cdots$        &[0.1-1.0]   &[0.6-0.8] &[0.6-1.0]     &$\cdots$     &[0.6-1.0]    &[0.5-0.7] &[0.5-0.7] & 0.3\\
   1.8  &   0.001   &   6    & $\le$\,2.1 &$\cdots$        &[0.3-2.0]   &0.8       &[0.7-1.2]     &[1.2-2.0]    &[0.7-0.8]    &[0.5-0.7] &[0.5-0.7] & [0.3-0.5]  \\
   1.8  &   0.001   &   10   & $\le$\,2.9 &$\cdots$        &[0.3-2.0]   &$\cdots$  &[0.7-1.0]     &[1.0-2.5]    &[0.7-1.0]    &[0.5-0.7] &[0.5-0.7] & 0.3\\
\hline
   1.8  &    0.01   &   2    & $\le$\,1.8 &$\cdots$        &[0.5-1.7] &$\cdots$ &[1.5-1.8]     &$\cdots$     &[1.7-1.8]    &[0.5-1.5] &0.5/[1.3-1.5] & [0.5-1.2]\\
   1.8  &    0.01   &   6    & $\le$\,5.0 &$\cdots$        &[1.2-5.0] &$\cdots$ &[1.2-3.3]     &[3.3-4.2]    &[2.0-3.3]    &[1.2-2.0] &[1.2-3.3]     & 1.5\\
   1.8  &    0.01   &   10   & $\le$\,9.4 &[5.0-9.0]       &[1.3-7.5] &[5.0-7.5]&[1.3-4.2]     &[2.5-8.3]    &[1.3-3.3]    &[1.3-2.0] &$\cdots$      & $\cdots$   \\
\hline
   1.8  &     0.1   &   2    & $\le$\,2.5 &$\cdots$        &[1.3-2.3]  &$\cdots$    &1.3/2.7       &$\cdots$      & 1.3         &1.3       & 2.3       & [1.7-2.0]   \\
   1.8  &     0.1   &   6    & $\le$\,8.6 &$\cdots$        &[3.3-8.3]  &8.3         &[3.3-7.5]     &3.3/[7.5-8.3] &[6.7-7.5]    &[4.2-5.8] & 4.2/5.8   & 5.0         \\
   1.8  &     0.1   &   10   & $>$\,15.0  & 5.0/15.0       &[6.7-15.0] &[13.0-15.0] &[6.7-13.3]    &[10.0-15.0]   &[6.7-11.7]   &[8.3-10.0]&[8.3-10.0] & $\cdots$    \\
\hline
\end{tabular}
\end{center}
    {\bf Notes:} All times (or epochs) are in days since the SN shock crosses $R_\ast$ in the \heracles\ simulation. ``N\,{\sc iii}--C\,{\sc iii}'' corresponds to the blend of multiplets due to both N\,{\sc iii} and C\,{\sc iii} and in particular N\,{\sc iii}\,4640.64\,\AA\ and C\,{\sc iii}\,4647.42\,\AA. ``$[a-b]$'' means that a given line is present between epochs $a$ and $b$. ``$a/b$'' means that the line was present at time $a$ and time $b$ but not in between. A single quoted epoch implies the corresponding line was present in only one of the computed spectra (and thus absent before and after).
\end{sidewaystable*}

\subsection{Lines of He, C, N, and O as diagnostics of the ionization evolution}
\label{sect_ionization}

Shock breakout from a RSG star is an extreme phenomenon that sees the exploded star change in surface temperature from a few 10$^3$ up to about 10$^5$\,K, though this jump is reduced for a higher CSM mass or extent \citep{moriya_rsg_csm_11,d17_13fs} -- this breakout phase also comes with a change in luminosity by a factor of $\sim$\,10$^5$, leading to a sudden and extended migration of the exploded star in the HR diagram. With our set of 27 different ejecta/CSM configurations, we obtain much diversity as evidenced by the range in peak luminosities, rise times, or time-integrated bolometric luminosity (see Table~\ref{tab_rhd_res} and Fig.~\ref{fig_rhd_res}). In this section, we turn to the corresponding spectral changes and in particular how emission lines during the IIn-phase vary in nature and strength -- the IIn phase corresponds to all times when there remains some optically-thick, unshocked, although potentially radiatively-accelerated CSM.

To make this analysis, we produced rectified spectra for all configurations and epochs (corresponding to a total of about 500 individual spectra). For each model set, we noted the epochs over which we could identify the presence of He\,{\sc i}\,5875.66\,\AA, He\,{\sc ii}\,4685.70\,\AA, C\,{\sc iii}\,5695.92\,\AA, C\,{\sc iv}\,5804.86\,\AA, ``N\,{\sc iii}--C\,{\sc iii}'' (this term corresponds here to the blend of multiplets due to both N\,{\sc iii} and C\,{\sc iii} and in particular N\,{\sc iii}\,4640.64\,\AA\ and C\,{\sc iii}\,4647.42\,\AA), N\,{\sc iv}\,7109.35\,\AA, N\,{\sc v}\,4612.72\,\AA, O\,{\sc v}\,5597\,\AA, and O\,{\sc vi}\,3820\,\AA. The results from this analysis are stored in Table~\ref{tab_spec_evol_sum}. These results have obvious limitations related to the fact that the spectral sequences only inform us on the actual times of the computed spectra. In particular, computations at the earliest times when the CSM is still cold are challenging so we make no claim on the presence or absence of any lines prior to the first spectrum of each sequence. Finally, we have tried to use a high-enough cadence to properly sample the spectral evolution. A higher cadence (meaning computing many more models and their spectra; doubling the cadence would mean computing 1000 models rather than 500) might reveal more subtle variations in some cases.

The first fundamental property of all simulations is the systematic rise and decline in ionization through the shock-breakout phase and beyond. This is easily explained from the preSN conditions characterized by a cool RSG atmosphere and CSM, the progressive heating and photoionization of this environment during breakout and shock propagation, and eventually the cooling and recombination of this material (CSM or shocked ejecta/CSM) through expansion and radiative losses. As long at the SN shock is deeply embedded in the optically-thick, unshocked CSM, the optical depth in the Lyman continuum and X-ray range is gigantic (i.e., the associated opacities are many orders of magnitude greater than that of electron scattering). Hence, photoionization during the IIn phase is entirely controlled by ``secondary photons'', that is photons emitted within the CSM. All ``primary photons'' from the SN shock are absorbed essentially on the spot. This photoionization is then caused by (quasi) blackbody radiation at an equilibrium temperature of a few $10^4$\,K and that gas or radiation temperature in the optically-thick CSM arises from the absorption of radiation and particles emitted by the shock (the gas properties from \heracles\ simulations of similar configurations are shown in \citealt{D15_2n,d17_13fs,dessart_wynn_23}).

\begin{figure*}
\centering
\includegraphics[width=0.8\hsize]{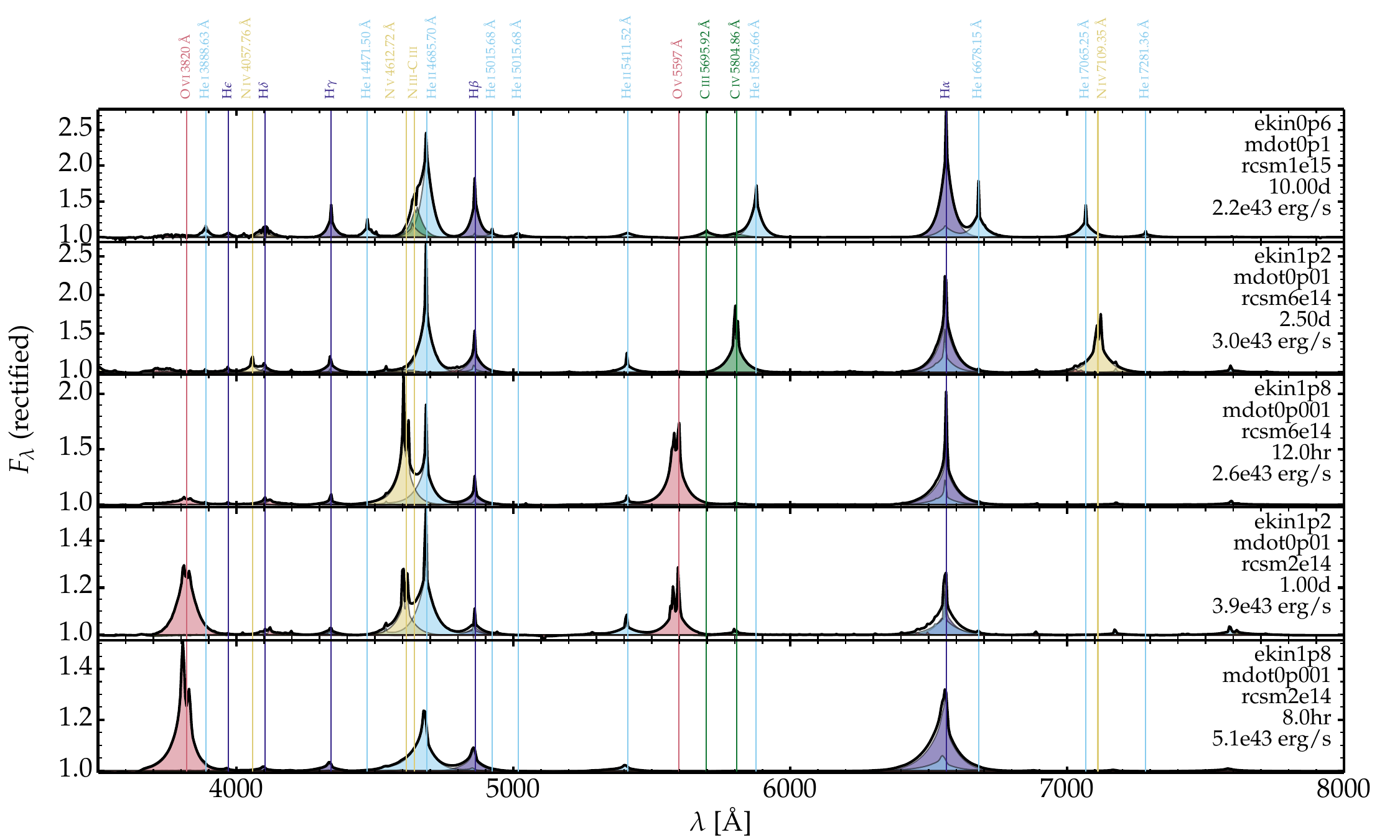}
\caption{Illustration of the spectral morphology during the IIn phase that results from a range of ionization of the unshocked CSM, from cool weakly ionized conditions at top to hot and highly ionized conditions at the bottom. The models shown here differ both in ejecta/CSM configuration and in post-breakout time. In all cases, He\two\,6560.1\,\AA\ is a contaminant to the H$\alpha$ emission.
\label{fig_spec_select}
}
\end{figure*}

Because of this rise and decline in ionization, the model spectra show an evolution of He lines with He\one\ $\rightarrow$ He\two\ $\rightarrow$ He\one\ (and eventually no He\one\ lines at all), C lines with  C\three\ $\rightarrow$ C\four\ $\rightarrow$  C\three\ (and eventually no C lines, at least in the optical), N lines with  N\three\ $\rightarrow$ N\four\ $\rightarrow$  N\five\ $\rightarrow$  N\four\ $\rightarrow$  N\three\ (and eventually no N lines at all). For O, the situation is more complicated since the models tend to predict lines only of O\five\ and O\six\ (i.e. O\,{\sc v}\,5597\,\AA\ and O\,{\sc vi}\,3820\,\AA), so only present when the ionization (or temperature) is high. To observe such a sequence in ionization (i.e., rise and decline) rather than a continuous decline requires an early discovery and prompt spectroscopic follow-up, something that was successfully achieved for SN\,2024ggi (see., e.g., \citealt{zhang_24ggi_24}) and obtained here in models rcsm6e14 and mdot0p01/mdot0p1. In some events characterized by a confined CSM, the surge in ionization is so short that only a decline may be observed (see, e.g., SN\,2013fs; \citealt{yaron_13fs_17}), as in models rcsm2e14 or mdot0p001 (i.e., computing models prior to this phase of high ionization is numerically challenging). At the opposite end (i.e., dense and very extended CSM), the ionization change through the shock-breakout phase is much more moderate with species remaining at most once or twice ionized (in the optical range). Finally, our simulations do not typically predict the simultaneous presence of lines (of a given element) from two consecutive ionization levels, such as N\three\ and N\four\ lines or O\five\ and O\six\ lines (there are some exceptions; e.g., model ekin1p2\_mdot0p01\_rcsm2e14 at 1\,d and shown in Fig.~\ref{fig_spec_select}). This arises mostly from the fact that although the unshocked CSM may have an ionization stratification, the bulk of the emission arises from a region of essentially uniform ionization. A natural way to break such a property is asymmetry, as may result from an asymmetric shock breakout or an asymmetric stellar environment (see, e.g., \citealt{gabler_3dsn_21,goldberg_sbo_22,vartanyan_ccsn_25}).

A related feature is that lines arising from ions that have a similar ionization potentials tend to be present at the same epochs. Again, this does not apply to H\one\ lines, which are present irrespective of the ionization level, thus in all model configurations and epochs. But He\one\ lines tend to be present along with N\three\ lines. He\two\ lines (just like H\one\ lines) are present over a wide range though of relatively high ionization (e.g., present together with N\three, N\four, or N\five\ lines). N\four\ lines tend to be present along with C\four\ lines, and for the same reason N\five\ lines tend to be present along with O\five\ lines (there are no C\five\ lines because of the huge associated ionization potential). Through our full set, the ion that is present only for the highest ionizations is O\six. In models with CSM, we find that the temperature/ionization of the unshocked CSM is never high enough to produce complete ionization so the optical spectra are never featureless (this occurs at the earliest post-breakout times in the No-CSM models).

Following Table~\ref{tab_spec_evol_sum}, we can discuss more specifically the spectral properties during the IIn phase in terms of explosion or CSM properties. The presence of He\one\ lines like He\,{\sc i}\,5875.66\,\AA\ is favored in models with dense and extended CSM (model mdot0p1 and rcsm1e15), or a low explosion energy -- He\one\ lines are present at later times when He\two\ disappears but this tends to occur in those cases during the CDS phase or after. He\two\ lines are present in all models at some stage but its subsistance requires high ionization so it vanishes more quickly in models with a lower explosion energy. In models with compact CSM (i.e., rcsm2e14), the ionization is high throughout the IIn phase in essentially all cases and thus electron-scattering broadened He\,{\sc ii}\,4685.70\,\AA\ is present (the weaker line He\,{\sc ii}\,5411.52\,\AA\ is also a useful diagnostic because it is isolated). C\,{\sc iii}\,5695.92\,\AA\ is typically rarely seen and quite weak, and instead all models predict strong C\,{\sc iv}\,5804.86\,\AA\ at some stage in the IIn phase, essentially for a subset of the epochs at which He\,{\sc ii}\,4685.70\,\AA\ is also predicted. N\three\ lines are mostly present in models with massive extended CSM (rcsm6e14 and rcsm1e15 or mdot0p1) and it tends to be present early on and later on following the rise and fall of the ionization. The strongest N\three\ line in the optical (i.e., N\,{\sc iii}\,4640.64\,\AA) produces an emission bump on the blue side of He\two\,4685.70\,\AA, but our models predict even more flux from the overlapping contribution due to C\,{\sc iii}\,4647.42\,\AA\ (this finding is not surprising given the presence of both N\,{\sc iv}\,7109.35\,\AA\ and C\,{\sc iv}\,5804.86\,\AA\ at some epochs and despite the overabundance of N relative to C). The models predict N\four\ lines unless the ionization remains too low (i.e., cases of dense or extended CSM). N\five\ and O\five\ lines require high gas ionization and thus shy away from the cases of dense and extended CSM. A higher kinetic energy boosts the CSM ionization and favors their presence. The most extreme of all is O\,{\sc vi}\,3820\,\AA, which is the most shortlived of all lines and requires the highest ionization, a situation only met for compact and tenuous CSM. Such a CSM tends to be optically thin sooner or strongly radiatively accelerated, so seeing an electron-scattering broadened O\,{\sc vi}\,3820\,\AA\ is challenging.

To summarize some of these properties, Fig.~\ref{fig_spec_select} presents a selection of spectra from different ejecta/CSM configurations and post-breakout epochs and stacked in order of increasing ionization from top to bottom. At top, a ``cool'' model spectrum shows the simultaneous presence of He\one, C\three, and N\three\ lines (model ekin0p6\_mdot0p1\_rcsm1e15 at 10\,d), followed by a ``warm'' model spectrum having He\two, C\four\ and N\four\ lines (model ekin1p2\_mdot0p01\_rcsm6e14 at 2.5\,d), then a ``warm/hot'' model spectrum with He\two, N\five\ and O\five\ lines (model ekin1p8\_mdot0p001\_rcsm6e14 at 12\,hr), and then two ``hot'' model spectra with simultaneously N\five, O\five, and O\six\ (model ekin1p2\_mdot0p01\_rcsm2e14 at 1\,d) or just O\six\ (model ekin1p8\_mdot0p001\_rcsm2e14 at 8\,hr). As written earlier, H\one\ lines (and He\two\ lines to a smaller extent) tend to be present over a wide range of ionization.

\begin{figure*}
\centering
\includegraphics[width=\hsize]{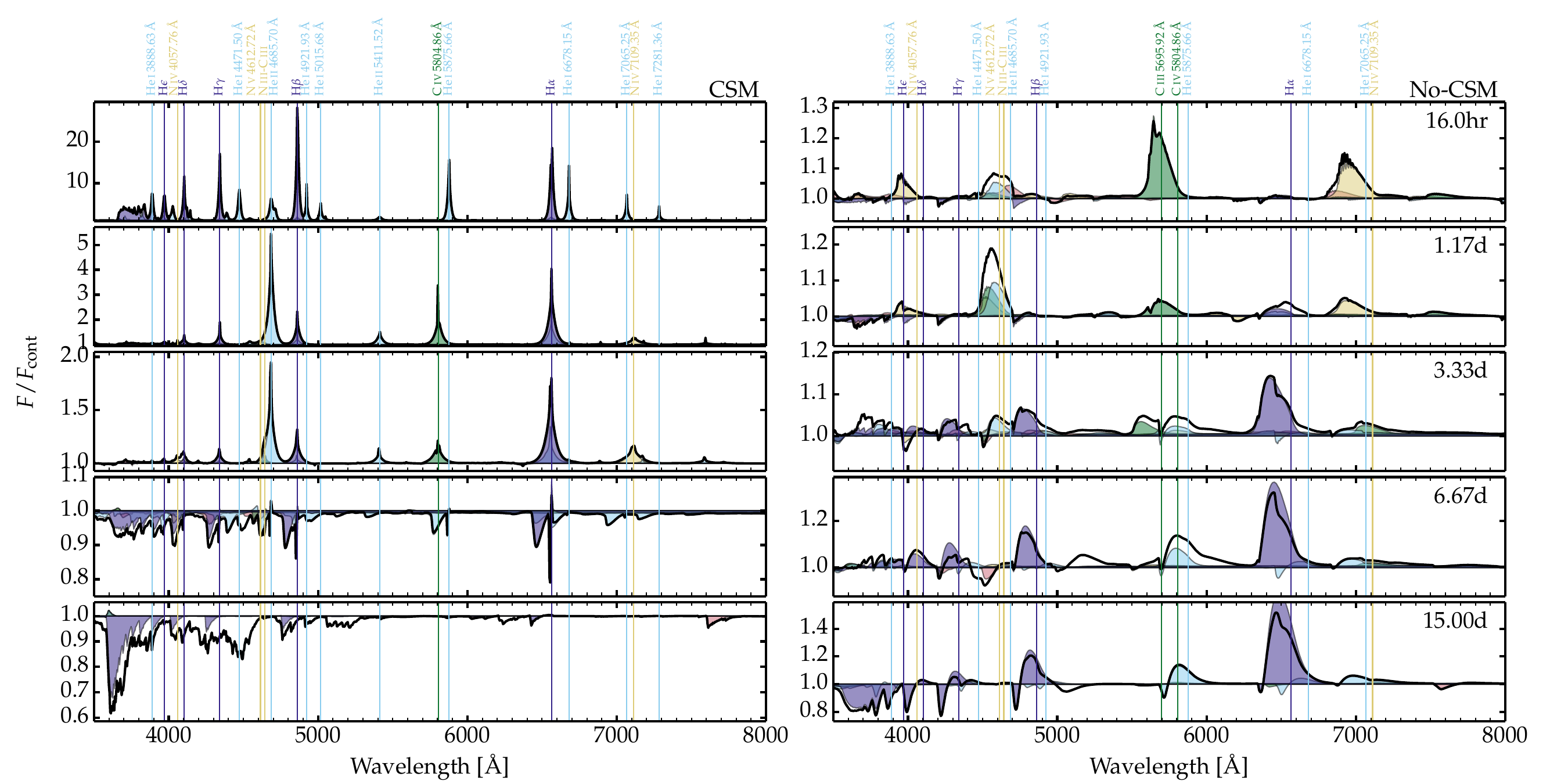}
\caption{Comparison of the spectral evolution of model ekin1p2 with CSM (left; case mdot0p01 and rcsm6e14) and without CSM (right). The rest wavelength of the main lines is indicated, which is also color-coded to differentiate different species. Lines affected by Doppler broadening exhibit a strong blueshift and may appear primarily in absorption (later epochs in the CSM case during the CDS phase) or primarily in emission (No-CSM case). (See Section~\ref{sect_csm_nocsm} for discussion.)
\label{fig_csm_nocsm}
}
\end{figure*}

\subsection{Comparison of results for configurations with and without CSM}
\label{sect_csm_nocsm}

In the preceding sections, we have discussed the impact of CSM on the dynamical and radiative properties of our Type II SN models. Corresponding results were confronted to the No-CSM case in Fig.~\ref{fig_rhd_res} (empty diamond symbols), as well as Figs.~\ref{fig_lbol_heracles}--\ref{fig_lc_uvw2_cmfgen} (dotted lines) and stored in Table.~\ref{tab_rhd_res}. Here, we discuss in detail the contrast in the associated spectroscopic properties.

Figure~\ref{fig_csm_nocsm} illustrates the fundamental change in spectral evolution and in particular in the level of ionization and line profile morphology when CSM is present and instead when the RSG star explodes in a vacuum (the CSM model used here is ekin1p2\_mdot0p01\_rcsm6e14 and the no-CSM counterpart just adopts a uniform wind mass loss rate of 10$^{-6}$\,\msunyr). These results were already shown and discussed in \citet{d17_13fs} with the models r1w1 (wind mass loss rate of 10$^{-6}$\,\msunyr) and for example r1w6 (high mass CSM density corresponding to wind mass loss rate of 10$^{-2}$\,\msunyr) but repeated here and reinforced with a more direct comparison -- still the present results are the same in nature as those of ten years ago.

The first important prediction is the presence of lines from ionized species in both cases. In fact, contrary to a common belief, the highest ionization is reached in the model without CSM, albeit short lived. For example, at 16\,hr, the No-CSM model exhibits lines of C\four\ and N\four, at a time when the model with CSM is cool and has mostly lines of He\one. As time goes on, the No-CSM model cools and exhibits lines of C\three\ and N\three, followed after a few days with lines mostly of H\one\ and He\one. In contrast, the model with CSM heats up during the first 1-2 days, switching from He\one\ to He\two\ lines etc, followed at later times after some cooling and recombination to a spectrum with He\one\ lines and H\one\ lines only, as in the No-CSM case. The different evolution followed by each model reflects the complex interplay between expansion and radiative cooling on the one hand, and photo-ionization heating on the other hand (for the CSM case). But clearly, the presence of He\two\,4685.70\,\AA\ is common to both CSM and No-CSM cases. This line has been observed for decades, and models without CSM could explain its presence at early times in SN\,2006bp or SN\,1999gi \citep{dessart_05cs_06bp}. Today, it is often referred to as a ``ledge'' feature (see, e.g., \citealt{hosseinzadeh_21yja_22}, \citealt{pearson_18lab_23}, \citealt{shreshta_23axu_24}).

The other difference, which is more striking, concerns line profile morphology. With CSM, the model exhibits IIn signatures as long as there is optically-thick unshocked CSM, followed by a phase during which the spectrum is quasi-featureless (i.e., when the photosphere is in the CDS). Over the 15\,d period shown, all lines evolve from having a narrow core with symmetric extended wings that are electron-scattering broadened to being Doppler-broadened with mostly blueshifted absorption. In contrast, without CSM, the lines are broad, Doppler-broadened and strongly blueshifted immediately at shock breakout as the progenitor surface is promptly accelerated after shock passage (the No-CSM model is in the ejecta phase at all epochs). This blueshift of all lines is an optical-depth effect and persists until the end of the photospheric phase, hence for several months in a typical Type II-Plateau SN \citep{DH05a,anderson_blueshift_14}.

The differences in profile morphology between the CSM and No-CSM cases are not limited to the initial 15\,d covered by Fig.~\ref{fig_csm_nocsm}. Models predict that differences persist at later times \citep{HD19}, with a bluer optical color, an excess H$\alpha$ emission with reduced absorption (see also, for example, \citealt{gutierrez_ha_14,gutierrez_pap1_17,gutierrez_pap2_17}). Additional signatures such as a UV excess may arise if weak power is injected in the outer ejecta by persisting interaction with the progenitor wind \citep{dessart_csm_22,dessart_late_23,wynn_pap2_24}, as observed in SN\,2023ixf \citep{bostroem_23ixf_24,bostroem_uv_25,wynn_sed_25}. Indeed, once the (outer) ejecta structure has been modified by interaction with CSM, it carries the imprint of that interaction for weeks, months and years, and typically only modified by further interaction with additional CSM at further distances.

\section{Conclusions}
\label{sect_conc}

We have presented radiation-hydrodynamics and radiative-transfer calculations of Type II SNe interacting with CSM with a focus on the first 15\,d after the SN shock crosses $R_\ast$. Compared to our previous work, we cover in a systematic way a broader range of explosion and CSM properties, specifically by modeling the explosion of a 15\,\msun\ star characterized by an ejecta kinetic energy of 0.6, 1.2, and $1.8 \times 10^{51}$\,erg, a CSM corresponding to a wind mass loss rate of 0.001, 0.01, and 0.1\,\msunyr, and a CSM extent of 2, 6, and $10 \times 10^{14}$\,cm. The results of this work are in line with \citet{d17_13fs} and \citet{dessart_wynn_23} but this new, extended, regular grid of models offers an unprecedented treasury for the analysis of future observations.

Compared to explosions in a vacuum (i.e., the No-CSM counterpart), the presence of CSM alters the dynamics and the escaping radiation from the SN by reducing the peak luminosity, broadening the light curve, and potentially extending the initial high-brightness phase through the extraction of ejecta kinetic energy. Consequently, decelerated ejecta material and swept-up CSM pile up into a dense shell, whose velocity is slightly smaller than the maximum ejecta velocity. We discussed how these various quantities relate to each other. A wide range of radiative acceleration of the unshocked CSM is also obtained.

We illustrated the importance of H$\alpha$ for inferring the dynamical evolution of the interaction, in part because H\one\ lines are present whatever the ionization. High-cadence monitoring, preferentially at high or moderate spectral resolution, can provide information on the size and density of the optically thick CSM during the IIn phase, the velocity and mass of the dense shell in the CDS phase, or the acceleration of the unshocked CSM located at and beyond the photosphere.

We also described the potential range in spectral diversity during the IIn phase, which is primarily driven by the variation in ionization of the unshocked CSM. These modulations in ionization are function of the post-breakout time, the CSM density and extent, as well as the explosion energy, such that much degeneracy prevails. With our model grid, we identified spectral properties reflecting a low ionization (He\one\ and N\three\ lines), a moderate ionization (He\two, C\four, and N\four\ lines), a high ionization (He\two, C\five, and N\five), and a very high ionization (He\two\ and O\six\ lines). We found that lines of C, N, and O are systematically present at the same time (e.g., C\,{\sc iv}\,5804.86\,\AA\ and N\,{\sc iv}\,7109.35\,\AA) because of the similarity in the corresponding ionization potential and despite the differences in abundances. We also found that N\,{\sc iii}\,4640.64\,\AA\ and C\,{\sc iii}\,4647.42\,\AA\ are both contributors to the narrow emission on the blueside of He\,{\sc ii}\,4685.70\,\AA.

Finally, we commented on the critical differences brought in with the presence of CSM. With CSM, the rise in ionization is delayed and the evolution of profile morphology exhibits at least three phases (IIn phase, CDS phase, and ejecta phase; see also \citealt{dessart_review_26}) -- the present simulations are mostly limited to the first two phases. During the IIn phase, line profiles all exhibit symmetric emission with a narrow core and extended electron-scattering broadened wings. During the CDS phase, the profiles are hybrid with a narrow P-cygni profile near line center from the distant unshocked CSM and a blueshifted absorption arising from the CDS. In contrast, without CSM, the rise in ionization is prompt and the profile morphology is immediately dominated by Doppler-broadened lines with a strong blueshift caused by optical-depth effects. In particular, He\,{\sc ii}\,4685.70\,\AA\ is present in both the CSM and No-CSM cases, but it is systematically broad and blueshifted in the No-CSM case.

The present work has a number of limitations. The radiative transfer technique uses the Sobolev approximation. We also adopt the temperature structure from \heracles\ in order to capture the temperature structure, controlled by the dynamics and the presence of the shock, but this temperature is compromised by the assumption of LTE for the gas (e.g. Saha ionization equilibrium) in \heracles. We also adopted a parametrized approach in which the ejecta kinetic energy, the CSM mass and the CSM extent are set without any global physical consistency. For example, the wind mass loss might be larger in higher mass progenitors whereas we used a fixed progenitor mass of 15\,\msun\ for all simulations. The wind mass loss may also correlate with the preSN mass, for example enhanced mass loss producing partially stripped progenitors -- the underlying assumption here was that whatever the adopted mass loss only negligible stripping results. Resolving these limitations will require code developments but also a much better understanding of mass loss in massive stars. Until then, a parametrized approach is the only practical way to proceed.

\begin{acknowledgements}

  LD acknowledges support from the ESO Scientific Visitor Program for a visit to ESO-Garching during the summer 2025. This work was granted access to the HPC resources of TGCC under the allocation 2024 -- A0170410554 on Irene-Rome made by GENCI, France. This research has made use of NASA's Astrophysics Data System Bibliographic Services.

\end{acknowledgements}


\onecolumn

\appendix

\section{Additional illustrations}
\label{sect_appendix}

\begin{figure*}[h]
\centering
\includegraphics[width=0.7\hsize]{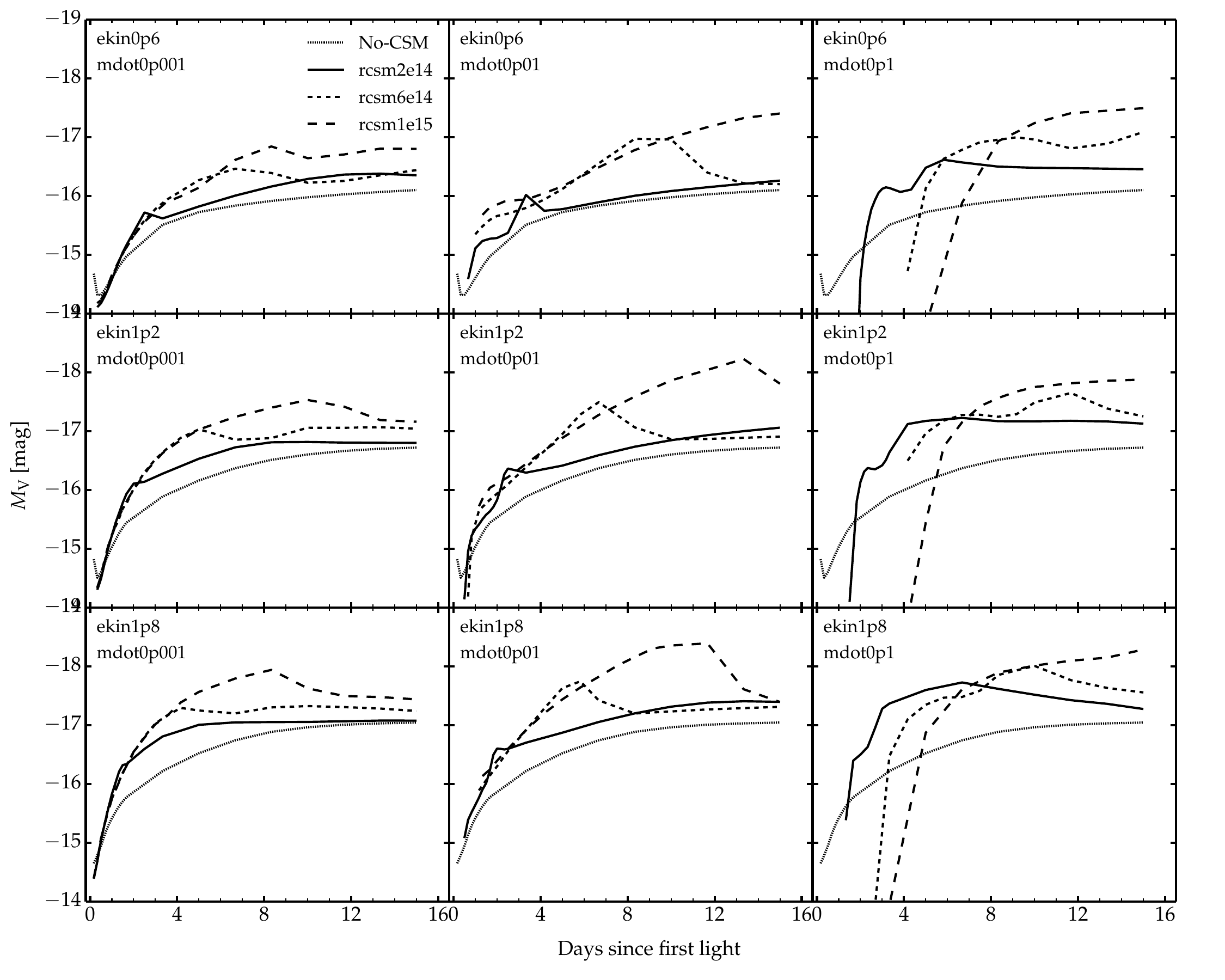}
\caption{Same as Fig.~\ref{fig_lc_uvw2_cmfgen} but now showing the $V$-band light curves computed with \cmfgen\ for our model set.
\label{fig_lc_v_cmfgen}
}
\end{figure*}

\begin{figure*}[h]
\centering
\includegraphics[width=0.7\hsize]{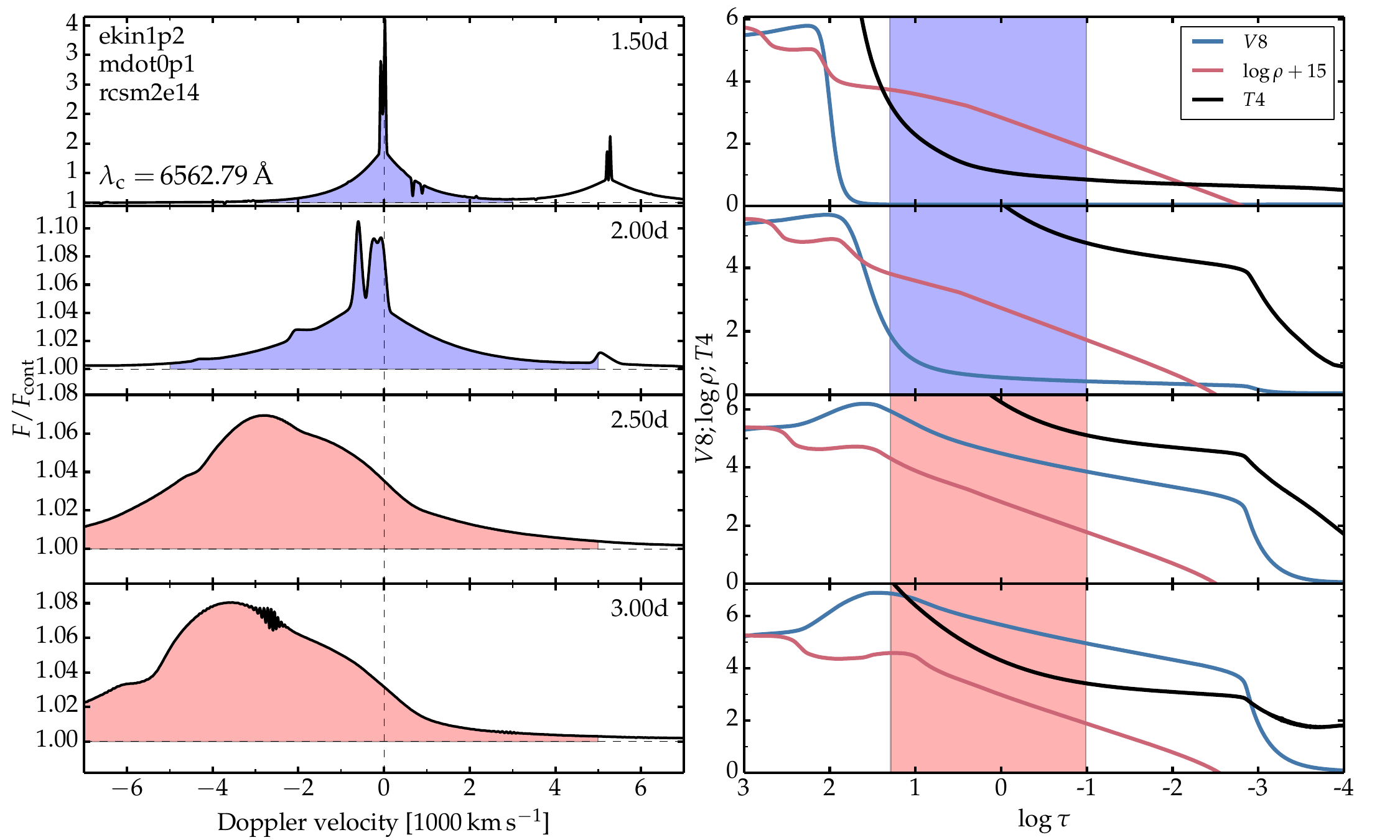}
\caption{Same as Fig.~\ref{fig_halpha_ekin1p2_mdot0p01_rcsm6e14} but for model ekin1p2\_mdot0p1\_rcsm2e14. All contributing lines are included in this figure, including He\one\ at 6678.15 and He\two\ at 6560.09\,\AA\ (this line dominates the emission seen at 2.0\,d and remains a strong contributor at 2.5 and 3.0\,d). This configuration with a more compact and denser CSM fosters a strong radiative acceleration of the unshocked CSM (causing the disappearance of the SN shock at all epochs shown), visible in the blue-shifted emission for the last two epochs (i.e., two bottom rows).
\label{fig_halpha_ekin1p2_mdot0p1_rcsm2e14}
}
\end{figure*}

In this section, we present additional illustrations to complement the main text. Figure~\ref{fig_lc_v_cmfgen} illustrates the $V$-band light curve for our model set as computed with \cmfgen\ (see Fig.~\ref{fig_lc_uvw2_cmfgen} for the $UVW2$-band countpart). We also show in Fig.~\ref{fig_halpha_ekin1p2_mdot0p1_rcsm2e14} the simultaneous  evolution of the H$\alpha$ region and the structure of the spectrum formation region in model ekin1p2\_mdot0p1\_rcsm2e14.

\end{document}